\documentclass{emulateapj}
\usepackage{apjfonts}
\usepackage{natbib}
\usepackage{epsfig,amssymb,amsmath}
\usepackage{color}

\newcommand{\ha}{H$\alpha$}
\newcommand{\sfre}{$\langle SFR \rangle_8$}

\begin{document}

\title{Measuring Galaxy Star Formation Rates From Integrated Photometry: Insights from Color-Magnitude Diagrams of Resolved Stars}

\author{Benjamin D. Johnson\altaffilmark{1},
Daniel R. Weisz\altaffilmark{2},
Julianne J. Dalcanton\altaffilmark{2},
L. C. Johnson\altaffilmark{2}, 
Daniel A. Dale\altaffilmark{3},
Andrew E. Dolphin\altaffilmark{4},
Armando Gil de Paz\altaffilmark{5},
Robert C. Kennicutt, Jr.\altaffilmark{6},
Janice C. Lee\altaffilmark{7},
Evan D. Skillman\altaffilmark{8},
Benjamin F. Williams\altaffilmark{2},
}

\altaffiltext{1}{Institute d'Astrophysique de Paris, CNRS, UPMC, 98bis Bd Arago, Paris 75014, France}
\altaffiltext{2}{Department of Astronomy, Box 351580,  University of Washington, Seattle, WA 98195, USA}
\altaffiltext{3}{Department of Physics and Astronomy, University of Wyoming, Laramie, WY 82071, USA}
\altaffiltext{4}{Raytheon, 1151 E. Hermans Road, Tucson, AZ 85756, USA}
\altaffiltext{5}{CEI Campus Moncloa, UCM-UPM, Departamento de Astrof\'isica y CC. de la Atm\'osfera, Facultad de CC. F\'isicas, Universidad Complutense de Madrid, Avda. Complutense s/n, 28040 Madrid, Spain}
\altaffiltext{6}{Institute of Astronomy, University of Cambridge, Madingley Rd, Cambridge CB3 0HA, UK}
\altaffiltext{7}{Space Telescope Science Institute, 3700 San Martin Drive, Baltimore, MD 21218, USA}
\altaffiltext{8}{Department of Astronomy, University of Minnesota, 116 Church St. SE, Minneapolis, MN 55455, USA}

\begin{abstract} 
We use empirical star formation histories (SFHs), measured from HST-based resolved star color-magnitude diagrams, as input into population synthesis codes to model the broadband spectral energy distributions (SEDs) of ~50 nearby dwarf galaxies ($6.5 < \log \mbox{M}/\mbox{M}_* < 8.5$, with metallicities $\sim 10$\% solar).  In the presence of realistic SFHs, we compare the modeled and observed SEDs from the ultraviolet (UV) through near-infrared (NIR) and assess the reliability of widely used UV-based star formation rate (SFR) indicators.  In the $FUV$ through $i$ bands, we find that the observed and modeled SEDs are in excellent agreement.  In the \emph{Spitzer} 3.6\micron~and 4.5\micron~bands, we find that modeled SEDs systematically over-predict observed luminosities by up to $\sim 0.2$ dex, depending on treatment of the TP-AGB stars in the synthesis models.   We assess the reliability of UV luminosity as a SFR indicator, in light of independently constrained SFHs. We find that fluctuations in the SFHs alone can cause factor of $\sim 2$ variations in the UV luminosities relative to the assumption of a constant SFH over the past 100 Myr.  These variations are not strongly correlated with UV-optical colors, implying that correcting UV-based SFRs for the effects of realistic SFHs is difficult using only the broadband SED.  Additionally, for this diverse sample of galaxies, we find that stars older than 100 Myr can contribute from $< 5$-100\% of the present day UV luminosity, highlighting the challenges in defining a characteristic star formation timescale associated with UV emission.  We do find a relationship between UV emission timescale and broadband UV-optical color, though it is different than predictions based on exponentially declining SFH models.  Our findings have significant implications for the comparison of UV-based SFRs across low-metallicity populations with diverse SFHs.
\end{abstract}


\section{Introduction}

Measuring the stellar mass and star formation rate (SFR) from a galaxy's observed spectral energy distribution (SED) relies on stellar population synthesis (SPS) models.  These models combine knowledge of stellar evolution and stellar spectra to convert between observations and physical quantities. In recent years it has become common to compare observed galaxies to model SEDs across a range of wavelengths, to derive multiple galaxy properties (e.g. stellar mass, SFR, metallicity) self-consistently. Fitting model SEDs to observed SEDs is now done for both low and high redshift galaxies \citep[e.g., ][]
{arnouts07, salim07, schaerer10, whitaker12, curtis-lake12, maraston12, mentuch12}, and from ultraviolet (UV) to far-infrared (FIR) wavelengths \citep[e.g., ][]{silva98, daCunha, noll09}.

Star formation histories (SFHs) are a critical component of SED modeling. A different SFH can change the relationship between physical quantities and the SED. However, the SFH of individual galaxies is usually poorly or only coarsely known, and some assumption about its form must then be made. The simplest models assume a constant SFR, to derive linear scalings between SFR and luminosity \citep[e.g., ][]{K98}.  More sophisticated modeling involves allowing the SFR to vary with time, though it is usually parameterized to be a smoothly varying function.  A common parametrization is the $\tau$-model, where the SFR declines exponentially \citep{tinsley68, madau98, kauffmann03a, sedfit}.  The timescale and amplitude of this parameterized SFH is then constrained by the SED itself.  

Difficulties in this approach arise from: 1) well-known and significant degeneracies between SFH, dust attenuation, and stellar metallicity \citep[e.g., ][ and references therein]{johnson07_fits, sedfit}; and 2) biases in the derived parameters due to true SFHs that deviate from the assumed parameterization (e.g., with a different long-term SFR evolution or variable SFR on short timescales). Recently \cite{lee10} have shown, using SFHs drawn from semi-analytic models, that determinations of the physical parameters of high-redshift galaxies can be significantly biased if they are derived from fitting SED models that assume a simplified or mismatched SFH \citep[see also][]{stringer11, pforr12}. These results depend on the adopted semi-analytic model of the SFH, and are valid at high-redshift.  How well they apply to real galaxies in the low-redshift universe is unknown. 

Instead of assuming a parameterized SFH, strong constraints on the real SFH can be obtained from a galaxy's resolved stellar populations \citep[e.g., ][]{tosi89, dolphin02}.  The location of individual stars in a color-magnitude diagram (CMD) constrains their evolutionary state, and can be used to infer the SFH for an assumed initial mass function (IMF). Studies of the resolved stellar populations of nearby galaxies are now routine with HST \citep[e.g., ][]{sanna09,angst,lcidV, grochalski12, dalcanton12}. With the SFH thus constrained by the CMD of resolved stars, we can determine the impact of realistic SFHs on typical conversions between observed and physical properties.  

One advantage of using the SFH measured for real galaxies, as opposed to a SFH drawn from semi-analytic models, is that it is possible to directly compare the measured SED to the SED inferred from the SFH. This comparison allows us to test the consistency of the SFH and population synthesis models with broadband observations from the ultraviolet (UV) to the near-infrared (NIR). Another advantage of using the SFH measured for real galaxies is that we can explore the effects of other galaxy properties on the SED (e.g., reddening by dust). The realistic SFH inferred from the CMD of individual stars serves to fix a large number of parameters in the SED model that usually have to be fit at the same time as other, possibly degenerate, parameters.

The paper is organized as follows. In \S\ref{sec:data} we describe the sample of galaxies, the broadband UV through MIR observations, the SFHs as derived from the resolved stellar CMDs of each galaxy, and how we predict the broadband SED using the CMD-constrained SFHs as input. In \S\ref{sec:compare} we compare the predicted luminosities to the observed luminosities in every band, highlighting discrepancies in the NIR. In \S\ref{sec:nir} we explore the possible origins of these discrepancies, including uncertainties in the population synthesis model ingredients. In \S\ref{sec:uv} we consider the effects of dust attenuation, metallicity, and stochastic sampling of the IMF on the observed SED. In \S\ref{sec:model_sfr} we investigate how the SFH affects the conversion of UV luminosity to SFR.

\section{Data}
\label{sec:data}

%%%%%%%%%%%%%%%%
%FIGURE - Sample
%%%%%%%%%%%%%%%
\begin{figure}
\epsscale{1.2}
\plotone{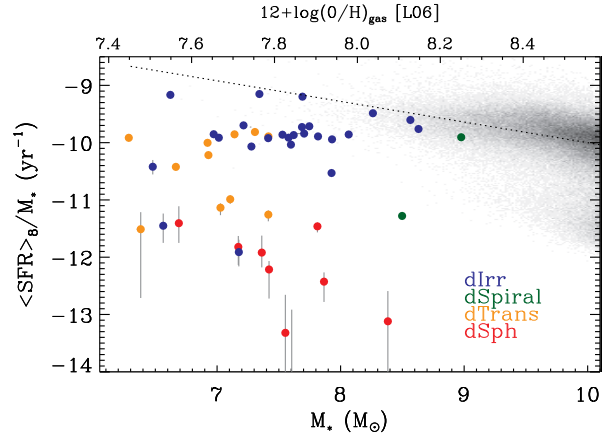}
\figcaption{Specific star formation rate versus stellar mass for the sample galaxies, where both have been derived from the SFH. For the stellar mass, we have applied a correction based on the amount of 3.6\micron~light falling outside the HST footprint to obtain an estimate of the total stellar mass.  The error bars are derived from the Monte Carlo realizations of the SFH, and encompass 68\% of the distribution of values.  For the stellar mass these errors are typically smaller than the symbol size. Each point is color-coded by the morphological type. The top axis shows the gas-phase metallicity inferred from the mass-metallicity relation of \cite{lee06_mz}. Grey-scale shows galaxies from the SDSS main galaxy sample.  The dahsed line is an extrapolation to lower mass of the SF sequence determined by \citet{schiminovich07}.
\label{fig:sample}}
\end{figure}

\subsection{Sample}
\label{sec:data:sample}

We analyze galaxies from the ACS Nearby Galaxy Survey Treasury \citep[ANGST; ][]{angst}, which consists primarily of dwarf galaxies (M$_* < 10^{9}$ M$_\sun$) with distances less than 4 Mpc. The sample spans a range of colors, morphologies (dE to dIm), and SFHs.  We have chosen a subset of ANGST galaxies for which HST observations sample a significant fraction of the total galaxy extent, and where the depth and quality of the observations provide for robust measures of the SFH \citep{weisz11_angst}.  The 49 selected galaxies are listed in Table \ref{tbl:pars}, along with their distances and other global properties.  In Figure \ref{fig:sample} we show the specific star formation rate (sSFR, the SFR divided by the stellar mass) versus stellar mass (where both quantities were derived from the CMDs, see \S\ref{sec:data:sfhs} and \S\ref{sec:model}) for the sample galaxies, color-coded by morphological type (\S\ref{sec:data:pars}).

\subsection{Broadband SEDs}
\label{sec:data:seds}
The integrated galaxy SEDs are derived from broadband imaging available from \emph{GALEX} \citep{galex}, SDSS, and \emph{Spitzer} \citep{spitzer}.  The \emph{GALEX}  and \emph{Spitzer} imaging has been obtained as part of the 11HUGS and LVL surveys respectively \citep{lee_uv, dale_lvl}, and we refer the reader to these papers for detailed discussion of the data reduction. Additional optical imaging for galaxies outside the SDSS footprint has been obtained by \cite{cook12}, through the Johnson-Cousins $UBVR$ filters. Unfortunately, the low surface brightness of these galaxies makes NIR $J,H$, and $K$ measurements difficult from the ground \citep[e.g., ][]{mcintosh06}, and we do not consider these bands here. 

We have measured the broadband luminosity that falls within the intersection between the HST footprint and large apertures designed to encompass the entire UV and NIR extent of the galaxy as follows. For the SDSS imaging we estimate the background from Gaussian fits to the lower 85\% of flux values in elliptical annuli extending from 1.5 to $\sim 2$ times the semi-major axis of the galaxy aperture. These estimates are consistent with the SDSS pipeline values outside the galaxy extent \citep[see ][for a discussion of biases in the SDSS background determination near large galaxies]{blanton_background}.  For the \emph{Spitzer} Infrared Array Camera \citep[IRAC,][]{irac} imaging we use the background determinations of \cite{dale_lvl}. For the \emph{GALEX} imaging we estimate the background by median filtering the \emph{GALEX} pipeline-produced background images in the same elliptical annuli as was used for the optical backgrounds.  The \emph{GALEX} pipeline-produced background images include masking of detected sources and account for the Poisson statistics of the low count-rate images. We subtract these backgrounds from the flux that falls in the HST footprints and the galaxy apertures.  

For \emph{GALEX} and \emph{Spitzer} we use the standard photometric zeropoints.  For the SDSS data the photometric calibration is taken from the calibration data provided for each SDSS imaging frame.  Masking of foreground stars and artifacts follows \cite{dale_lvl} and \cite{lee_uv}, though we have inspected the foreground masks by hand for these galaxies to be sure that no UV bright clusters are mistakenly masked.  For several galaxies the presence of nearby and extremely bright foreground stars makes accurate photometry impossible. These galaxies are DDO78, KKR025, and IKN.  Though we do not consider the photometry at any wavelength for these galaxies, we retain them in the sample since they can still contribute to conclusions based solely on the modeled fluxes (see \S\ref{sec:model} and \S\ref{sec:model_sfr}).  The quoted photometric uncertainties of the \emph{GALEX} and \emph{Spitzer} fluxes are typically dominated by calibration uncertainty \citep{dale_lvl, lee_uv}, but systematics related to sky background estimation (especially in the ground-based $u$, $U$, and $z$ bands) and unsubtracted foreground stars likely contribute substantially to the true photometric uncertainty.

Our photometry is reported in Table \ref{tbl:photom}.  The fraction of the total $g$ or $B$-band flux that falls within the HST footprint is listed in Table \ref{tbl:pars}. We also list in Table \ref{tbl:pars} the estimated fraction of the blue optical light that is contributed by resolved stars used in the SFH determination (\S\ref{sec:data:sfhs}).  Corrections for Milky Way reddening are derived from the reddening maps of \cite{SFD}; the adopted conversions from E(B-V) to $A_\lambda$ are given in Table \ref{tbl:reddening}.  For the \emph{GALEX} bands these conversions follow \cite{galex_ngs} and for the optical we derive the conversions from the Milky Way extinction curve of \cite{CCM} with R$_V=3.1$ .  We convert between absolute magnitude ($M_{\lambda_{e}}$) and solar luminosities using 
\begin{equation}
\log \nu L_\nu = \log c/\lambda_e +(51.595-M_{\lambda_e})/2.5 -\log L_\sun
\end{equation}
 where $\lambda_e$ are the effective wavelengths of the bands\footnote{We use the definition of \cite{schneider83} for $\lambda_e$.} as given in Table \ref{tbl:reddening} and $L_\sun=3.827\times10^{33}$ erg/s.

\subsection{Additional Galaxy Properties}
\label{sec:data:pars}
We also consider measures of internal dust attenuation and metallicity, which are known to affect the SED.  Accurate estimates of dust attenuation are notoriously difficult to obtain. For the vast majority of the sample, measurements of the Balmer decrement are not available, and so attenuations cannot be measured in this way \citep{lee09_had}. Such measurements also only sample the very youngest stars, not the attenuation of the stellar population as a whole. However, the broad wavelength coverage of this sample allows us to use the infrared-to-UV ratio (IRX), which is available for the majority of the sample. Following \cite{hao11}, we estimate the FUV attenuation in magnitudes from the ratio of 24\micron~to FUV flux as 
\begin{equation}
A_{fuv}=2.5\log \lgroup 1+ \eta {\nu L_\nu(24\micron) \over \nu L_\nu (FUV)} \rgroup
\end{equation}
where $\eta=3.89$ is a scaling factor determined from observations.  This relation has been calibrated by \cite{hao11} using attenuations derived from the Balmer decrement, albeit for a sample of galaxies with higher luminosities and larger average dust attenuation. For galaxies undetected at 24\micron~(24 of the 49 sample galaxies), we use the 1$\sigma$ 24\micron~flux limit to define an upper limit on the attenuation.  The resulting FUV attenuations are given in Table \ref{tbl:pars}.  For the majority of the sample, A$_{fuv}<0.35$ mag, consistent with their low luminosities. Therefore, the uncertainties in the method used to derive the attenuation are of minimal importance, because the absolute attenuations are undoubtedly small.

Morphological T-types are taken from \cite{rc3} and \cite{karachentsev04} and are given in Table \ref{tbl:pars}.  Following \cite{weisz11_angst} we define the galaxy types dwarf spheroidal (dSph, T$\le 0$), dwarf irregular (dIrr, T$=10$), and dwarf spiral (dSpiral, $7<T<10$). We include the dwarf transition (dTrans) type, defined as galaxies with detectable gas content but undetectable \ha~emission. The galaxy metallicity may be estimated from a mass-metallicity or luminosity-metallicity relation. We use the 4.5\micron\ luminosity-metallicity relation of \cite{berg12} to estimate the gas-phase metallicity.  The resulting estimates of the gas-phase metallicity are listed in Table \ref{tbl:pars}.  The metallicities are typically well below solar. 

%%%%%%%%%%%%%
%TABLE: wavelengths and reddening
%%%%%%%%%%%%%
\begin{deluxetable}{lcc}
\tablecolumns{3}
\tablecaption{Filters \label{tbl:reddening}}
\tablehead{\colhead{Band} & \colhead{$\lambda_{eff}$} & \colhead{$A_\lambda/E(B-V)_{MW}$} \\
\colhead{} & \colhead{(\AA)} & \colhead{}}
\startdata
$        FUV$ &   1528.1 &      7.9 \\
$        NUV$ &   2271.1 &      8.1 \\
$          u$ &   3546.0 &      5.0 \\
$          U$ &   3571.2 &      5.0 \\
$          B$ &   4344.1 &      4.2 \\
$          g$ &   4669.6 &      3.8 \\
$          V$ &   5455.6 &      3.2 \\
$          r$ &   6156.2 &      2.8 \\
$          R$ &   6441.6 &      2.6 \\
$          i$ &   7471.6 &      2.1 \\
$          I$ &   7993.8 &      1.9 \\
$          z$ &   8917.4 &      1.5 \\
$ 3.6\micron$ &  35416.6 &      0.0 \\
$ 4.5\micron$ &  44826.2 &      0.0 \\
$ 5.6\micron$ &  56457.2 &      0.0 \\
$   8\micron$ &  78264.8 &      0.0 \\
\enddata
\end{deluxetable}

\subsection{Deriving SFHs from Resolved Stars} 
\label{sec:data:sfhs}
We derived SFHs for the sample from their optical stellar CMDs.  The methodology involves matching the observed density of stars in color-magnitude space to linear combinations of the CMD density expected from simple stellar populations (SSPs) of various ages, including reddening by dust and observational effects modelled with extensive artificial star tests; a more detailed description can be found in \cite{weisz11_angst}. The CMD fitting takes into account Poisson statistics when measuring fit quality, and thus implicitly includes the effects of stochasticity in the population of the IMF on the number of stars of a given mass (or luminosity) in the error estimate (see \ref{sec:model:uncertainties}).   The constraint on the most recent SFH comes primarily from main sequence stars, whereas the SFH between $50-500$ Myr is largely constrained by easily identified helium burning (HeB) stars, which follow a rough luminosity-age relation \citep{dohm_palmer}.  

The CMDs were derived with stellar metallicity in each age bin as a free parameter, and assuming a Salpeter IMF from 0.1 to 100 M$_\sun$ (slightly different from the IMF used in \cite{weisz11_angst}) and the stellar evolutionary tracks of \cite{girardi00}. Differential extinction of young stars is included following the model of \citet{dolphin03}.  In this model a flat distribution of extinction values is applied to all stars younger than 100 Myr, with the maximum extinction increasing linearly from $A_V=0.0$ at an age of 100 Myr to $A_V=0.5$ mag at an age of 40 Myr.  By increasing the width of the main sequence in the simulated CMDs this differential extinction model gives vastly improved fits to the data over no extinction. In \S\ref{sec:dust} we discuss the impact of differential reddening and independent estimates of dust content in more detail.   The temporal resolution of the derived SFHs is $\Delta \log t=0.1$, which is coarser than the time resolution used for the SFH derivation in \citet{weisz11_angst}, though finer than the time resolution displayed in that work. The SFHs are derived within the range 4 Myr$<\log t_{lookback}< 14.1$Gyr.  An example SFH is shown in Figure \ref{fig:sfh_example}.

We have used the SFHs to derive \sfre, the average SFR over the last 100 Myr, and M$_*$, the current stellar mass, by integrating the SFH over time, accounting for stellar death using population synthesis models (see \S\ref{sec:model}). The resulting properties are listed in Table \ref{tbl:pars}. The metallicities inferred from the CMD fitting, while uncertain, are consistent with the gas-phase metallicities of \S\ref{sec:data:pars} and Table \ref{tbl:pars}.

\subsection{Predicting Integrated Luminosities from the SFH}
\label{sec:model}
We derive the expected luminosity for each galaxy within the area covered by the HST footprint, by inputting the measured SFHs into a population synthesis code \citep[see ][for an early example of this method]{wyder01}. To be consistent with the derivation of the SFH we adopt the Flexible Stellar Population Synthesis \citep [FSPS,][]{conroy09, conroy10_III} model (version 2.3, revision 60) as our fiducial model.  This code uses the same \citet{girardi00} stellar evolutionary tracks for the main sequence up to the AGB phase.  The default FSPS models use the AGB star isochrones of \citet{marigo08}, with modifications as suggested in \citet{conroy10_III}.  To maintain consistency with the derivation of the SFH from resolved stars, we have used FSPS \emph{without} these modifications; the effects of changing the AGB isochrones on the predicted SED will be discussed in \S\ref{sec:agb}.  Post-AGB evolution utilizes the tracks of \cite{VW94}.  For main sequence and giant stars, the stellar spectra in FSPS models are from BaSeL 3.1 \citep{basel3.1}.  Spectra of TP-AGB stars are from the empirical library of \citet{LM02}, with extensions redward of 2.5\micron~using \citet{aringer09} for carbon-rich TP-AGB stars and the PHOENIX `BT-SETTL' spectral library for oxygen-rich TP-AGB stars.  The spectra of post-AGB stars are from \citet{rauch02} and the spectra of OB and Wolf-Rayet stars are from \cite{smith02}.   We use a Salpeter IMF from 0.1 to 100 M$_\sun$ for consistency with the CMD fitting. We consider a fiducial metallicity of 0.2$Z_\sun$ for all models. This value is consistent with the measured gas-phase metallicities, the metallicities expected from the mass-metallicity relation, and estimates of the metallicity from the CMD fitting. The metallicity has only a  modest effect on the UV and optical luminosities, as we discuss further in \S\ref{sec:metal}, and we do not expect significant effects from limiting the models to a single metallicity. These models do not include dust attenuation or emission, except for the circumstellar envelopes of AGB stars.  The stellar synthesis models implicitly assume that the CMD is fully sampled, and thus neglect stochastic effects due to small numbers of stars at high stellar masses or rare evolutionary phases.

We have also considered the population synthesis models of \cite{CB07}, which are based on the `Padova 1994' \citep{bertelli94} stellar evolutionary tracks, supplemented by the \cite{marigo08} evolutionary tracks for TP-AGB stars. The main difference between \cite{girardi00} and `Padova 1994' is a warmer (and bluer)  giant branch in the former \citep{BC03}.  There are some differences between the FSPS models and \cite{CB07}  in the stellar spectral libraries used. However, we find that using the \cite{CB07} models instead of our fiducial models does not significantly affect the predicted SED (except in the \emph{Spitzer} 5.8 and 8\micron~bands, see \S\ref{sec:nir}), and does not change our conclusions.

%%%%%%%%%%%%%%%%
%FIGURE - SFH example
%%%%%%%%%%%%%%%
\begin{figure}
\epsscale{1.2}
\plotone{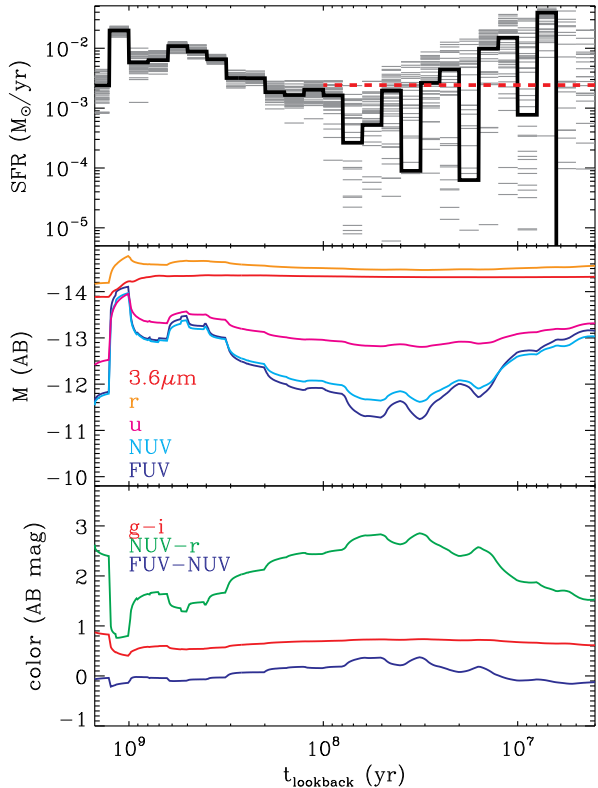}
\figcaption{{\it Top:} An example star formation history.  The black histogram shows the best-fit SFH determined from fits to the optical stellar CMD for NGC4163, at higher temporal resolution than shown in \cite{weisz11_angst}. The grey histograms show the MC realizations of the SFH. The red dashed line shows the SFR averaged over the past 100 Myr ($\langle SFR \rangle_8$). {\it Middle:} An example of the modeled luminosity evolution. The colored lines give the absolute AB restframe magnitude in several bands as a function of lookback time, expected given the SFH in the top panel. 
 {\it Bottom:}  Modeled color evolution for the same galaxy.
\label{fig:sfh_example}}
\end{figure}

In detail, we generate SPS models having constant SFR (of 1 M$_\sun/$yr) from zero age to the duration of each of the time bins used in the SFH reconstruction, and zero SFR thereafter. The resulting spectra of the SPS models (with each model corresponding to a single time bin in the SFH) are then interpolated logarithmically in time to a set of ages $t-t_i$ where $t_i$ is the beginning time of the bin, and the appropriate normalization is applied. 
\begin{equation}
\ell_\lambda^{mod}(t)=\sum_i^{N_{bin}} SFR_i \hat{\ell}_{i,\lambda}(t-t_i) e^{-\tau_\lambda(t-t_i)}
\end{equation}
where $SFR_i$ is the SFR in time bin $i$ from the CMD based SFH, $\hat{\ell}_{i,\lambda}(t')$ is the luminosity at wavelength $\lambda$ of the truncated constant SFR SPS model at age $t'$, and $\tau_\lambda(t')$ is the dust extinction optical depth at wavelength $\lambda$ towards stars of age $t'$.  In order to match the reddening model used when deriving the SFH from the CMD (\S\ref{sec:data:sfhs}) the maximum of $\tau_\lambda(t-t_i)$ is taken to linearly decrease from $\tau_V=0.5$ for $t-t_i \leq 40$ Myr to $\tau_V =0$ at $t-t_i  =100$ Myr.  To account for the distribution of extinctions toward stars of a given age in the Sextans A differential extinction model \citep[\S\ref{sec:data:sfhs} and][]{dolphin02} the spectrum of each bin, $\hat{\ell}_{i,\lambda}(t')$, is divided into a large number of equal pieces, each piece is extinguished using a different $\tau_V$, uniformly distributed up to the maximum for that time bin, and the pieces are then summed.  We note that this makes little difference from simply assuming one average value of $\tau_V$ per bin, regardless of the number of pieces. As the sample is primarily composed of low-mass, low-metallicity dwarf galaxies, the shape of the extinction curve is taken to be that of the SMC \citep{pei92}, and we neglect scattering.  The resulting $\ell_\lambda^{mod}(t)$ are convolved with the appropriate filter transmission curves to determine the broadband luminosity $L_\lambda^{mod}(t)$, which we refer to as the modeled (with dust) luminosity.  We also determine the modeled (intrinsic, dust-free) luminosity in all bands, denoted $L_{0,\lambda}^{mod}(t)$ hereafter, by setting $\tau_\lambda=0$ for all ages.  Stellar masses, accounting for stellar evolutionary effects, are derived in a similar way by replacing $\hat{\ell}_{i,\lambda}(t')$ with $\hat{m}_{i}(t')$, the stellar mass of all surviving stars and stellar remnants at age $t'$.  Typically this stellar mass is $\lesssim0.15$ dex smaller than the total stellar mass formed over the lifetime of the galaxy.

%%%%%%%%%%%%%%%%
%FIGURE - SED vs Bin
%%%%%%%%%%%%%%%
\begin{figure*}
\plotone{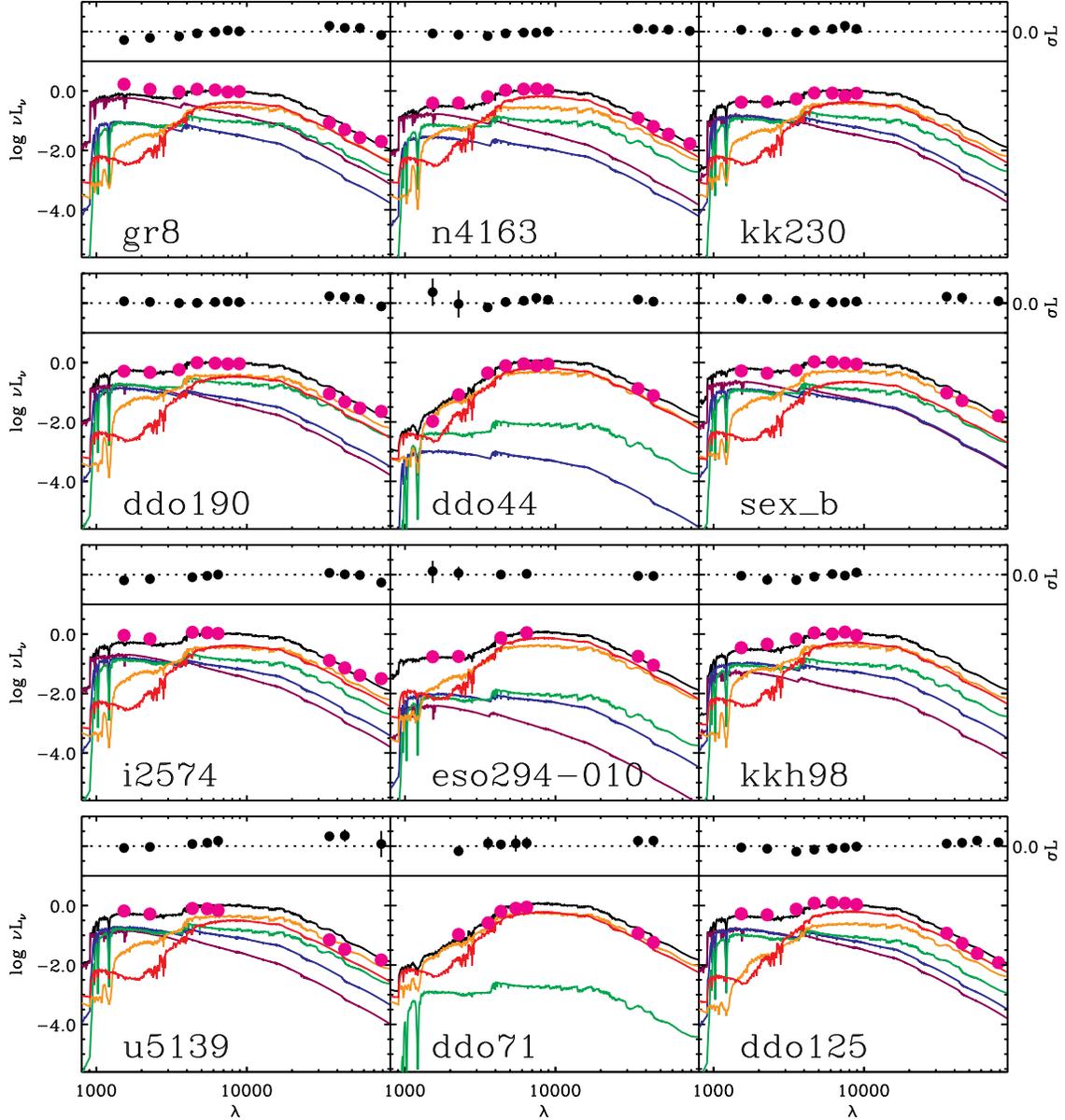}
\figcaption{The modeled and observed SED for 12 of the sample galaxies.  In each panel, the magenta points show the observed SED for one galaxy, while the black line shows the modeled $t_{lookback}=0$ SED.  The colored lines give the contribution to the present day spectrum of stars formed in different time bins: $6.7<\log t_l\le 7.3$ (purple), $7.3<\log t_l\le 8.0$ (blue), $8.0<\log t_l\le 8.7$ (green), $8.7<\log t_l\le 9.4$ (orange), $9.4 <\log t_l\le 10.15$ (red).  All SEDs are normalized by the 5500\AA~ luminosity of the modeled $t_{lookback}=0$ SED. At the top of each panel the black points show the residuals between the modeled and observed broadband luminosity ($\delta$L$=\log L_{mod}/L_{obs}$), with error bars giving the quadrature sum of the photometric error and the dispersion of the modeled flux from the different MC realizations.  The residuals are on the same scale as the SEDs.
\label{fig:evolsed}}
\end{figure*}

The SFR and duration of the most recent time bin is altered so that the total mass of stars formed in that bin is distributed over the interval 0 to 4 Myr.  This is because the derivation of the SFR from the CMD results in stars with ages $<4$ Myr being `assigned'  (or fit by stars with) the minimum isochrone age ($4\mbox{Myr}<t_{lookback}<5$Myr). We include an estimate the effects of nebular emission on the broadband luminosities, which we find to be less than 0.05 dex in all bands.

Figure \ref{fig:sfh_example} shows an example of the derived luminosity evolution of one galaxy.  It is clear that the discrete binning of the SFH has strong effects on the flux evolution, especially at large lookback time when the temporal resolution is longer than the lifetime of the dominant stars.  However, the fluctuations in SFR at large lookback times do not significantly affect the present day luminosities. At small lookback times the temporal resolution is better, and the effect of the binning on even the UV luminosity evolution is small. The effects of binning in the SFH are discussed further in \S\ref{sec:model:uncertainties}. 

\subsection{Uncertainties in the SFH}
\label{sec:model:uncertainties}
There are several sources of uncertainty in the measured SFHs that can affect the predicted luminosities.  The first source is random uncertainties due to the number of stars in each region of the CMD. Statistical uncertainties on the SFHs are computed through 50 Monte Carlo (MC) realizations.  For each MC test, a Poisson random noise generator is used to randomly resample the best fitting model CMD.  As a result, the uncertainties on the SFHs account for fluctuations in the number of stars used to derive the SFH in a given time bin.  For each galaxy we repeat the analysis described above for different MC realizations and then compute the dispersion in the resulting flux at each wavelength to obtain an estimate of the uncertainty of the predicted flux at the present day. This procedure also takes into account the covariance between adjacent bins. The typical uncertainty in the predicted FUV flux is $\sim 0.05$ dex, and is unbiased with respect to the FUV luminosity of the best-fitting SFH.  This uncertainty decreases to $\sim 0.01$ dex redward of the $u$ band.  A similar procedure is used to estimate the uncertainties in \sfre~and M$_*$ as derived from the SFH. Again, because of significant covariance between time bins the uncertainties on time averaged properties, including broadband luminosities, are significantly smaller than implied by quadrature sums of the uncertainties on the SFR in individual bins.

The second source of uncertainties in the SFHs are systematic effects due to differences between stellar models \citep[see, e.g., ][]{dolphin12, charlot96, conroy09, weisz11_angst}. Estimating the effect of these systematic uncertainties on the predicted fluxes requires deriving the SFH with different stellar models and then predicting the broadband flux with the same model. The differences in the predicted fluxes when using these different stellar models then give an estimate of the systematic uncertainty of the predicted flux. While \cite{weisz11_angst} have provided an estimate of the uncertainty in the SFH induced by uncertainties in stellar isochrones, a fair estimate of the resulting uncertainty in the integrated broadband luminosity requires predicting the integrated SED with the same (randomly) modified isochrones. This is beyond the scope of the current paper, but deserves treatment in future analyses. Readers are cautioned that the uncertainties in the physical parameters listed in Table \ref{tbl:pars} do not include the (often dominant) systematic uncertainties due to changes in the stellar models \citep{weisz11_angst, dolphin12}.

The third source of uncertainty in the predicted fluxes is the discrete time binning of the SFHs. At one extreme, the SFR may fluctuate more smoothly than the derived SFH, simply due to the binning in time of a smoothly rising or falling SFH.  At the other limit, the true SFR may vary on timescales shorter than the width of a given bin \citep[see, e.g., ][ for a discussion of the impact of SFRs that vary strongly within a given bin]{eskew11}. In general, these limitations will be a significant effect only for the wavelengths that are sensitive to timescales shorter than the resolution of the SFH.  For example, the ionizing flux ($\lambda<912$\AA) is nearly always sensitive to the SFR on timescales shorter than the temporal resolution of the SFH, and therefore significant artifacts of the binning scheme would be visible in the \ha~flux evolution, even at small lookback times. For the broadband wavelengths, however, the temporal resolution is sufficient for the prediction of present day luminosities.

\section{Comparison to Observed SED: Average Differences}
\label{sec:compare}

In this section we compare the modeled luminosities to the observed luminosities.  This comparison allows us to 1) verify the reliability of the SFHs, 2) explore the effects of additional galaxy properties on the SED, free from the potentially degenerate effects of SFH, and 3) test the predictions of population synthesis models in the UV and NIR.  

In Figure \ref{fig:evolsed} we show the modeled present-day SED for a subset of the galaxy sample, as well as the observed broadband luminosities.  We also show the contribution of stars formed in several different broad age bins to the present day SED, and the residuals between the modeled luminosities and the observed luminosities.  In general there is good agreement.

\begin{figure*}
\plotone{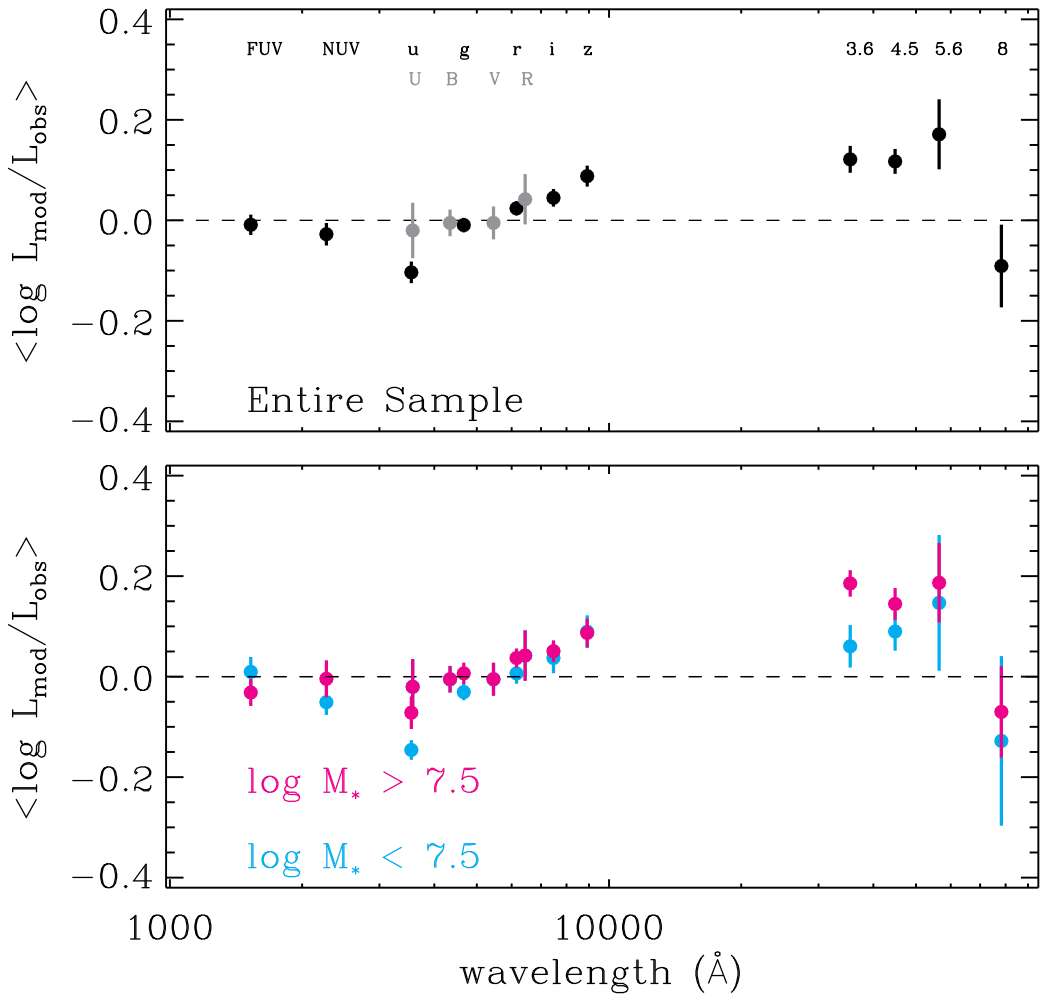}
\figcaption{Comparison of observed luminosities to modeled luminosities. \textit{Top:} Average differences between the absolute magnitudes measured from broadband imaging (and corrected for MW extinction) and the absolute magnitudes determined from the CMD-derived SFH and the FSPS models, as function of wavelength for the entire sample. The error bars give the standard error of the mean. Numbers below each point give the number of galaxies with flux differences available at that wavelength. The Johnsons-Cousins filters are shown in grey. \textit{Bottom:} The same, but with the sample split by stellar mass, showing that the offsets are not a strong function of stellar mass or properties that correlate with stellar mass.
\label{fig:avg}}
\end{figure*}

In Figure \ref{fig:avg} we plot the average differences between the observed luminosities and the modeled luminosities as a function of wavelength, with error bars showing the standard error of the mean. The first thing to note is the excellent agreement in the UV through optical ($FUV$ through $i$), where the overall normalization appears to be correct to within 0.1 dex, and often much better. The stars used for the SFH determination typically contribute $\sim$40\% of the total luminosity in the $g$ through $V$ bands (though this fraction varies significantly) suggesting that the SFH is correctly inferring the behavior of the remaining 60\% of the flux. 

However, in the NIR portions of the spectrum, we see significant systematic differences. Real galaxies appear to be fainter than the models in the \emph{Spitzer} IRAC 3.6, 4.5 and 5.8\micron~bands by an average of $\sim 0.2$ dex.  As discussed in \cite{conroy09} and \cite{mancone12}, the emission in this region of the SED is often dominated by more poorly understood phases of stellar evolution (particularly the TP-AGB phase), and it is here that different population synthesis models disagree most strongly.  This uncertainty is in contrast to the optical, where the stellar spectra and the evolutionary tracks of the dominant populations are more certain and do not vary as widely between authors. The origin of these systematic offsets will be discussed further in \S\ref{sec:nir}.  Here we note that splitting the sample by stellar mass, which correlates with many other parameters (e.g. metallicity, SFR) yields similar results for galaxies both more and less massive than $10^{7.5}$M$_\sun$

In Table \ref{tbl:offsets} we present, for each wavelength, the average and r.m.s. dispersion of the ratio of the modeled luminosity to the observed luminosity,  $\log (\mbox{L}_{mod}/\mbox{L}_{obs})$. We will refer to these ratios in each band as `offsets' or `excesses'.  In the optical the scatter is low ($\sim 0.1$ dex), and in all bands is less than 0.2 dex.

%%%%%%%%%%%%%%%%
%FIGURE - Average delta flux
%%%%%%%%%%%%%%%

\section{Differences in the NIR}
\label{sec:nir}
In this section we explore the differences between modeled and observed luminosity in the NIR that were found in \S\ref{sec:compare}.  The light in the NIR IRAC bands comes from a number of sources.  At 3.6 and 4.5\micron, the IRAC flux is dominated by the long-wavelength tail of cool luminous AGB, RGB, and RHeB stars, with a modest contribution from interstellar lines and dust. In the longer wavelength 5.6 and 8\micron~IRAC filters, the flux is increasingly dominated by emission from dust, especially PAH emission.  

The average differences between the modeled and observed luminosities in the stellar dominated 3.6\micron~and 4.5\micron~IRAC bands are important to consider in light of the widespread use of NIR photometry for stellar mass determinations \citep[e.g., ][]{lee06_mz, berg12, meidt12}, and ongoing uncertainty in the treatment of TP-AGB stars in population synthesis models. In the 3.6, 4.5, and 5.6\micron~bands, and to a lesser extent in the $i$ band, there is a significant offset, in the sense that the observed NIR luminosities are fainter than predicted by our fiducial model. There are several possible reasons for these offsets, which may also contribute to the scatter in the ratio of modeled to observed luminosity.  First, systematic uncertainties in the ancient SFH (that are not included in the MC-derived uncertainties, \S\ref{sec:data}) may cause offsets in the NIR bands.  The ancient SFH is most strongly constrained by the fainter, redder RGB stars, but we find no trend of the NIR offsets with the depth of the HST CMD; a thorough exploration of this possibility will require deeper CMDs (as are available for local group galaxies or the Magellanic clouds).  The second, more likely, explanation is related to the treatment of uncertain phases of stellar evolution in population synthesis models, especially the TP-AGB phase.  This is discussed in more detail below.

%%%%%%%%%%%%%%%%
%FIGURE - Average delta flux with modified agbs
%%%%%%%%%%%%%%%
\begin{figure}
\epsscale{1.2}
\plotone{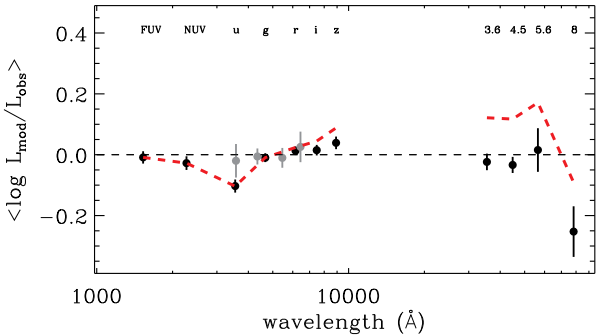}
\figcaption{Comparison of observed luminosities to modeled luminosities with reduced contribution to the NIR from TP-AGB stars, as suggested by \cite{conroy10_III}. Symbols are as in Figure \ref{fig:avg}.  The red dashed lines shows the results of Figure \ref{fig:avg}, where the TP-AGB contribution is not reduced. The systematic offsets in the NIR have been substantially reduced. 
\label{fig:agb}}
\end{figure}

\subsection{The Impact of AGB Star Prescriptions}
\label{sec:agb}
The contribution of TP-AGB stars to the NIR spectrum of galaxies is the subject of much debate \citep{maraston05, CB07, conroy09, kriek10, girardi10, zibetti12}.  Both the evolutionary paths and the NIR spectra of these stars are poorly constrained, and thus we are not surprised to find significant disagreements between the predicted and observed NIR luminosity.  To explore possible origins of the observed offsets, we have considered population synthesis models with different treatments of the TP-AGB phase. While this breaks the consistency between the population synthesis modeling and the derivation of the SFH, tests by \citet{girardi10} and \citet{melbourne12} have found that the CMD-based SFH of many of our sample galaxies are relatively insensitive to the treatment of the AGB phase, and produce nearly identical SFHs even when AGB stars are excluded from the fit.  This is because there are few of them relative to the RGB, and so they carry less weight in the SFH determination.

We first examine the effect of modifying the TP-AGB isochrones from the \cite{marigo08} treatment, as suggested by \cite{conroy10_III}.  This modification was made to better match the optical/NIR colors of Magellanic cloud globular clusters at the ages dominated by TP-AGB stars.  The net effect of this change is to significantly reduce the contribution of TP-AGB stars to the NIR SED.  Using the FSPS models with modified AGB isochrones to predict the SED as in \S\ref{sec:model}, we compare the newly predicted luminosities to the observed luminosities in Figure \ref{fig:agb}.  This figure can be directly compared to Figure \ref{fig:avg}.  Even though these galaxies are not post-starbursts, nor young enough that the NIR SED is dominated by TP-AGB stars \citep[see, e.g., ][]{kriek10, zibetti12}, there is a significant effect on the predicted IRAC luminosities such that the (average) agreement with the observed luminosities is improved.  The median ratio of the predicted to observed  luminosities at 3.6 and 4.5\micron~has decreased from 0.22 and 0.15 dex to 0.02 and -0.03 dex, respectively. This change in predicted luminosities also demonstrates that modeled NIR mass-to-light ratios are sensitive to these different treatments of the evolution of TP-AGB stars at least at the 0.2 dex level. 

In the \emph{Spitzer} Near-IR bands, we found that the modeled SEDs consistently overpredict the 3.6, 4.5, and 5.8\micron~luminosities by $\sim 0.2$ dex on average.  However, if the evolution of TP-AGB stars is altered from that of \cite{marigo08} then it is possible to obtain much better agreement.  This conclusion is qualitatively consistent with \cite{melbourne12}, who found that the \cite{marigo08} models tend to significantly overpredict (by a factor of two or more) the number and total luminosity of luminous TP-AGB stars that are present in the HST $H$-band imaging of many of these same galaxies.   When using the \cite{girardi10} evolutionary tracks for TP-AGB stars, \citet{melbourne12} found that the numbers of TP-AGB stars were better matched to observations, though the total TP-AGB luminosity was still overpredicted by factors of approximately two.  Fortunately, the fraction of the total $H$-band light due to TP-AGB stars is relatively small for these galaxies, and because this overprediction of TP-AGB luminosity is largely offset by an underproduction of RHeB luminosity in the models relative to the data, \citet{melbourne12} found the effect of this systematic overprediction of TP-AGB luminosity on the total $H$-band luminosity to be small. However, at 3.6 and 4.5\micron, where the fraction of the total luminosity due to TP-AGB stars is likely to be larger than at 1.6\micron~\citep[e.g., ][]{CB07}, and the fraction due to RHeBs smaller than at 1.6\micron, an overestimate of the TP-AGB luminosity may have larger consequences for the total luminosity than at $H$ band.  A detailed comparison of the predictions of different evolutionary tracks and spectral libraries will be the subject of a future work.

With a decrease in the contribution of TP-AGB stars to the integrated NIR SED, the observed 8\micron~luminosities are larger than the modeled 8\micron~luminosities, consistent with a small contribution of un-modeled dust in this band.  However, when the galaxies are split by stellar mass (or alternatively metallicity) as in the bottom panel of Figure \ref{fig:avg}, the low-mass, low-metallicity galaxies still show an apparent deficit of modeled 8\micron~luminosity. Since these galaxies are expected to have a smaller abundance of PAH grains and lower PAH emission \cite[e.g., ][]{madden00, hogg05_pah,marble10, wu11}, this deficit of the modeled 8\micron~luminosities may not necessarily be due to the lack of modeled dust emission from the ISM, but instead to the uncertain spectra of AGB stars at these wavelengths \citep[perhaps related to the un-modeled circumstellar dust around these stars, e.g., ][]{srinivasan11}.

\section{Impact of Metallicity, Stochasticity, and Dust Attenuation}
\label{sec:uv}

\subsection{Metallicity}
\label{sec:metal}

\begin{figure}
\epsscale{1.2}
\plotone{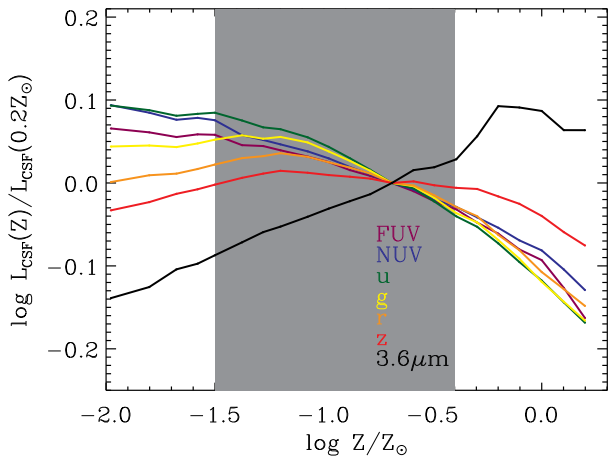}
\figcaption{The luminosity in several broad bands of a constant SFR model as a function of stellar metallicity, relative to the $Z=0.2Z_{\sun}$ models assumed in \S\ref{sec:model}. The grey shaded region indicates the plausible range of current gas-phase metallicity of our sample galaxies, inferred from emission lines and the NIR luminosity-metallicity relation.
\label{fig:met}}
\end{figure}

In \S\ref{sec:model} we fixed the stellar metallicity at 0.2Z$_\sun$, due to uncertainties in the stellar metallicities of the sample galaxies.  In this section, we examine the effect of variations of the true metallicity from this assumed value on the modeled luminosities.  We do this using a constant SFR as input to FSPS, and determine the SED for a variety of assumed metallicities.  The ratio of the resulting luminosity to the luminosity of the $Z=0.2Z_\sun$ model as a function of $Z$ over the plausible range of metallicities for our low mass sample of galaxies is shown in Figure \ref{fig:met}.

For the lower metallicity galaxies in our sample our assumed metallcity may lead to underestimates of the true luminosity by up to $\sim 0.1$ dex in the UV.  Conversely, the IRAC luminosities of low metallicity galaxies may be overestimated by as much as much as 0.1 dex, or more if the stellar metallicities of the stars contributing most strongly in these bands are significantly lower than the gas-phase metallicity. However, there are few galaxies at the very lowest metallicities, and tests assuming $Z=0.1Z_\sun$ for the stellar metallicity indicate that the average offsets change by less than 0.05 dex (and that $\langle L_{mod}/L_{obs}\rangle$ \emph{increases} slightly in the IRAC NIR bands).  Combined with the agreement seen in Figure \ref{fig:avg},  this suggests that the effect of stellar metallicity on the comparisons is minimal, though metallicity variations may contribute a small amount ($\lesssim 0.05$ dex) to the scatter of individual offsets at a given wavelength.

%%%%%%%%%%%%%%%%
%FIGURE - metallicity
%%%%%%%%%%%%%%%

\subsection{Stochastic Sampling of the IMF}
\label{sec:uv:stochastic}
At very low SFR, or alternatively for a small mass of `young' stars, it is possible that stochastic sampling of the IMF and the cluster mass function can lead to an apparent deficit of massive stars, due to their relative rarity \citep{cervino03, cervino09}.  This effect has been shown to be important in the interpretation of the H$\alpha$ flux of dwarf galaxies, which arises ultimately from the hydrogen ionizing emission of $\gtrsim 20$M$_\sun$ O and early B stars \citep{lee09_had, fumagalli11, eldridge12}.  This stochastic sampling of the IMF may also affect the UV flux of galaxies with low SFR. However, since the UV emission arises from more numerous, lower mass stars, the effect will always be less important than for H$\alpha$.

Because we have used population synthesis models that assume a fully sampled IMF when constructing the predicted SEDs and UV fluxes, our models do not include the scatter and bias in flux that is caused by stochasticity.  On the other hand, the uncertainties on the SFH derived from the CMDs do take this stochasticity into account implicitly in the fitting process, and generally rely on more numerous stars of lower masses.  Thus, if stochastic effects are important for the UV luminosity, we would expect a difference between the observed and predicted SED that is larger than the uncertainty in the predicted UV flux inferred from the MC realizations (in the sense that the observed SED is fainter than predicted on average). We would also expect an increased uncertainty or scatter in the predicted UV flux if stochastic effects were to be included. These differences would anti-correlate with the SFR of the galaxy, or more properly with the mass of stars formed over some characteristic UV emitting lifetime, because at higher SFRs there should be more complete sampling of the upper IMF.  

\begin{figure}
\epsscale{1.2}
\plotone{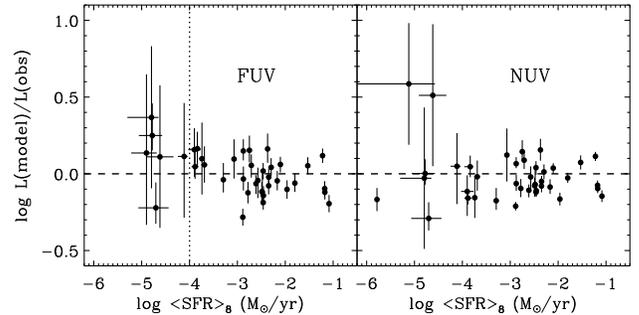}
\figcaption{\sfre, the star formation rate averaged over 100 Myr inferred from HST CMDs, is plotted against the difference between the observed and modeled fluxes at FUV ($left$) and NUV ($right$). The dotted line in the left-hand panel marks the SFR below which stochastic effects might be expected to become important. Vertical dashed lines mark perfect agreement between modeled and observed luminosities. 
\label{fig:stoch}}
\end{figure}

The contribution of stochasticity to the luminosity differences is explored indirectly in Figure \ref{fig:stoch}, where we plot the FUV and NUV luminosity differences vs \sfre.  While the scatter in luminosity differences appears to increase slightly at lower \sfre, we find no clear correlation, suggesting that stochastic effects do not play a large role in the observed UV luminosities of these galaxies. While a proper accounting of the effect of stochasticity on the average and scatter in UV luminosity would require modeling the drawing of individual stars from the IMF and cluster mass function given a total mass of stars formed in the various time intervals \citep[][ though note that the SFHs used in our study implicitly include sampling of the cluster mass function]{SLUG}, the comparisons shown in Figure \ref{fig:stoch} are sufficient to rule out the stochasticity as a major contributor to individual offsets between the observed and modeled UV luminosities.

%%%%%%%%%%%%%%%%
%FIGURE - stochasticity
%%%%%%%%%%%%%%%

\subsection{Dust Attenuation}
\label{sec:dust}

In \S\ref{sec:model} we modeled the luminosities both including the effects of the differential reddening model that is required to match the observed optical CMD and without this differential reddening (i.e., the intrinsic, dust-free luminosity).  The reddened luminosities were found to provide a good match to the observed luminosities in the UV through optical bands.  In this section, we compare the extinctions inferred from from the reddened and unreddened model luminosities to commonly-used independent estimates of the dust attenuation based on the observed ratio of the IR to UV luminosity.

\begin{figure}
\plotone{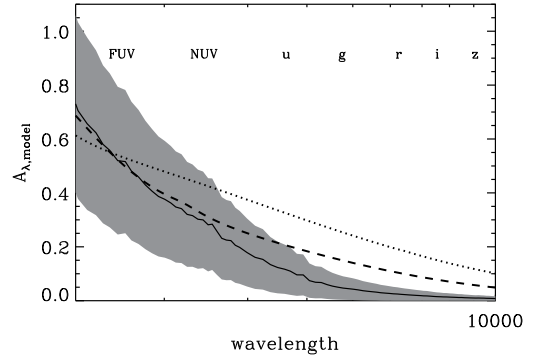}
\figcaption{The average model effective extinction as a function of wavelength for the sample ({\it solid line}).  The gray shaded region shows the r.m.s. scatter in the extinction among the galaxies (due to variations of the SFH in combination with the differential extinction model).  For comparison, a \citet{calzetti00} attenuation curve and the \citet{pei92} SMC extinction curve are shown as the dotted and dashed lines respectively. Both curves have been normalized by the average $A_{fuv,mod}$.
\label{fig:a_model}}
\end{figure}

The effective total extinction at any wavelength can be derived from the model luminosities of \S\ref{sec:model} by 
\begin{equation}
A_{model, \lambda}=-2.5 \log \lgroup \frac{L_{0,\lambda}^{mod}}{L_{\lambda}^{mod}} \rgroup
\end{equation}
where $L_{0,\lambda}^{mod}$ is the intrinsic model luminosity derived from the SFH without extinction by dust and $L_{\lambda}^{mod}$ is the luminosity derived including both SFH and the differential extinction model. The average and scatter of these modeled attenuations for the sample are shown as a function of wavelength in Figure \ref{fig:a_model}.  Note that, because the reddening only affects stars younger than 100 Myr, and because we have assumed the relatively steep attenuation curve of the SMC, the extinction of bands redward of NUV are small.  Also, because the differential extinction model is fixed, the scatter is due entirely to the different SFHs among the sample galaxies.

\begin{figure}
\plotone{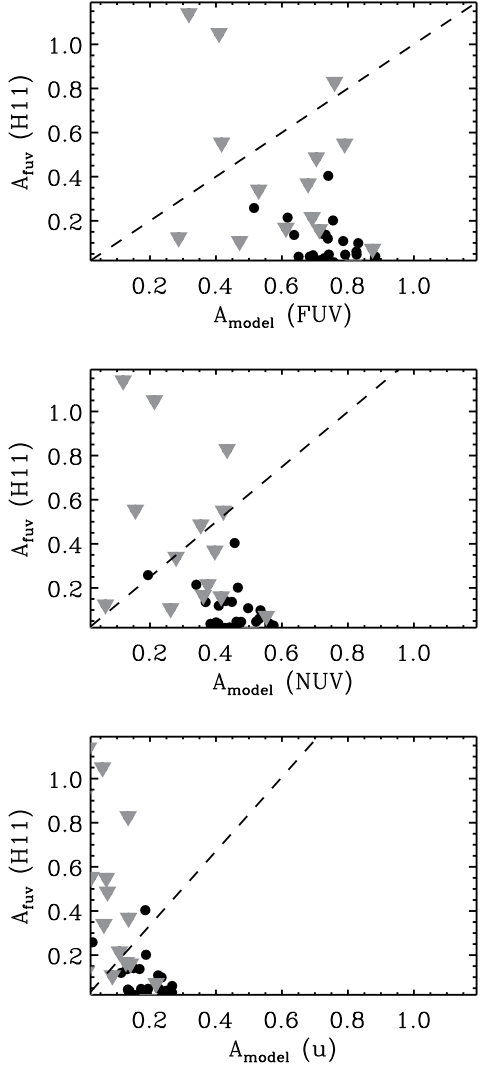}
\figcaption{Comparison of the model effective extinction in the FUV, NUV, and $u$ bands to $A_{fuv}$ estimated from the ratio of 24\micron\ to FUV luminosity following \citet{hao11}. Gray triangles indicate upper limits to $A_{fuv}$ based on 24\micron\ upper limits. Dashed lines are the expected relation for a \citet{calzetti00} attenuation curve.  
\label{fig:dust}}
\end{figure}

In Figure \ref{fig:dust} we compare $A_{model}$ for the FUV, NUV, and $u$ bands to the FUV attenuation $A_{fuv}$  inferred from the ratio of 24\micron\ to FUV luminosity based on the calibration of \citet{hao11} (\S\ref{sec:data:pars}) In nearly every case $A_{fuv}$ is significantly smaller than the attenuation derived from the CMD based SFH and differential extinction model.  However, the calibration of $A_{fuv}$ given by \citet{hao11} is largely based on much more massive and dusty galaxies, where older stars may contribute a larger fraction of the dust luminosity, and where the effective attenuation curve is significantly shallower or flatter than we have derived in Figure \ref{fig:a_model}.

\begin{figure}
\plotone{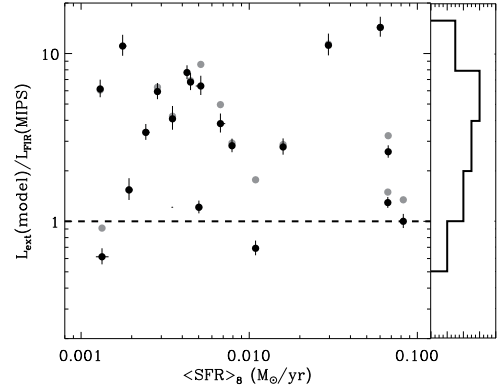}
\figcaption{Comparison of the observed total FIR luminosity, calculated from MIPS luminosities following \citet{dale_lvl}, to the total extinguished luminosity predicted given the extinction model.  Vertical error bars only include photometric uncertainties.  Grey points include a correction for the fraction of young star light falling outside the HST aperture. The histogram in the right panel indicates the number of galaxies in logarithmic bins of the ratio of the extinguished model luminosity to the observed IR luminosity. 
\label{fig:lir}}
\end{figure}

A more direct comparison of the modeled extinctions to the infrared luminosity can be made by integrating the difference between the spectra modeled with and without differential extinction (the $\ell_\lambda^{mod}$ of \S\ref{sec:model}) and assuming that this extinguished luminosity is reradiated in the infrared.  In Figure \ref{fig:lir} we compare this extinguished luminosity to the IR luminosity derived from {\it Spitzer} MIPS observations by \citet{dale_lvl}, both with and without corrections for the smaller aperture of the HST data.  This comparison is only possible for the 20 galaxies in the sample with detections in all MIPS bands.

We find that the total extinguished model luminosity is larger than the observed IR luminosity by a factor of 4 on average.  While this may be a result of modeled extinctions that are larger than the true extinction, tests have shown that assuming zero extinction results in very poor fits to the optical CMD.  Alternatively, the assumption that all extinguished light is reradiated in the IR is likely to be incorrect. Scattering by dust will serve to decrease the amount of extinguished light in our model SEDs that is ultimately absorbed by dust \citep[e.g.][]{WG00}. However, because the scattered light is primarily at UV wavelengths, a significant contribution of scattered light would then result in a poor match between the observed and modeled UV luminosities in Figure \ref{fig:avg}.  Assuming the attenuation curves of \citet{WG00} for a clumpy, cloudy geometry with an SMC extinction curve and a V band extinction of 0.25 mag (as in the differential extinction model of \S\ref{sec:model}) yields absorbed IR luminosities that are consistent with the observed IR luminosities on average (though with large scatter), but also results in modeled FUV luminosities that are brighter than the observed luminosity by $\sim 0.2$ dex on average. 

It is possible that some moderate amount of differential extinction, combined with the relatively gray attenuation curves of \citet{WG00}, might adequately match the optical CMD without predicting too much IR luminosity and that corresponding changes in the recent SFH would maintain the agreement we find between the observed and modeled UV and optical SEDs. A detailed exploration of this possibility would benefit greatly from multi-band stellar photometry and CMDs to further constrain the differential extinction, as are being obtained in M31 as part of the Panchromatic Hubble Andromeda Treasury survey \citep[PHAT, ][]{dalcanton12}.

%%%%%%%%%%%%%%%%
%FIGURE - dust A model
%%%%%%%%%%%%%%%

%%%%%%%%%%%%%%%%
%FIGURE - dust A_fuv comparison
%%%%%%%%%%%%%%%

%%%%%%%%%%%%%%%%
%FIGURE - L_IR comparison
%%%%%%%%%%%%%%%

\subsubsection{IRX-$\beta$}

The intrinsic UV color of galaxies has implications for measurement of dust attenuation via restframe UV colors using the so-called IRX-$\beta$ diagram \citep{MHC99}, which relates the IR to UV luminosity ratio (IRX, a measure of attenuation) to the observed UV spectral slope $\beta$ (where $f_\lambda\propto \lambda^{-\beta}$).  Over the last decade, there has been increasing evidence that normal galaxies are shifted to redder UV colors in this diagram than starburst galaxies \citep[e.g., ][]{seibert05, kong04, johnson07_irx, galex_ngs, dale07, dale_lvl, boquien12}, with significant implications for the measurement of attenuation in large samples of high-redshift galaxies \citep[e.g., ][]{reddy12, smit12_hizsfrd}.  It is often proposed that variations in the SFH are the cause for this scatter, since more passive galaxies may have a redder intrinsic UV color. 

\begin{figure}
\plotone{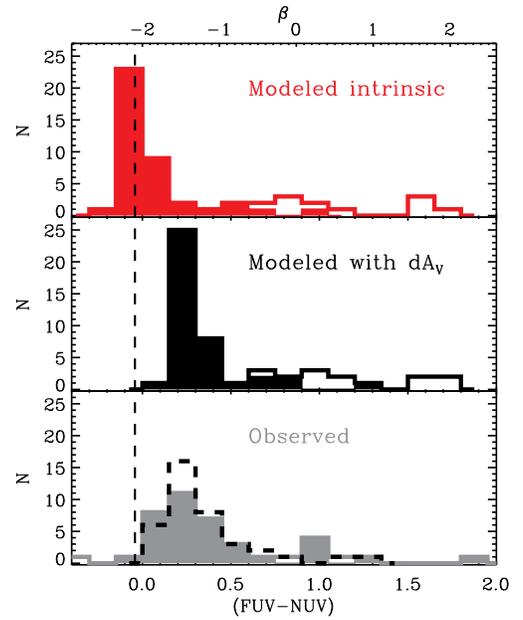}
\epsscale{1.3}
\figcaption{The distribution of FUV-NUV color for the sample galaxies.  {\it Top:} The intrinsic FUV-NUV color derived from the luminosities modeled without differential attenuation.  The filled histogram includes all sample galaxies not classified as dwarf spheroidals.  {\it Middle:} As for the top, but for the modeled FUV-NUV color including differential extinction.  {\it Bottom:} The observed FUV-NUV color distribution for all galaxies with detections in both bands is shown in grey.  The filled histogram is for only the galaxies that are not dwarf spheroidals. The dashed black histogram shows the distribution of modeled FUV -NUV colors including differential extinction and a contribution from photometric and model uncertainties in the FUV-NUV color.  In all panels the vertical dashed line marks the intrinsic, dust-free color for a constant SFR model.  The top axis indicates the UV spectral slope $\beta$, estimated as $\beta=2.3(FUV-NUV)-2$.
\label{fig:beta}}
\end{figure}

Using the SFHs derived in \S\ref{sec:model} we explore this possibility.  This sample is particularly well suited to such a study as many of the sample galaxies have low IRX but have been found to be shifted to redder UV colors than starburst galaxies \citep{dale_lvl}.   First, we determine the intrinsic FUV-NUV color implied by the SFHs of \S\ref{sec:data:sfhs}; that is, the color that is obtained without application of the differential reddening model.  The distribution of these intrinsic UV colors are shown as the red histogram in Figure \ref{fig:beta}, where the filled histogram does not include the galaxies classified as dwarf spheroidals.  We find a small scatter of these intrinsic colors around the value predicted for constant SFR.  

We have also determined the distribution of UV colors implied by the SFHs and the differential reddening model, shown as the black histogram in Figure \ref{fig:beta}, finding that differential reddening shifts the galaxies to redder UV colors (by $~0.3$ mag or 0.6 in $\beta$) but without significantly increasing the scatter in color.  Finally, we construct the distribution of observed UV colors (the grey histogram in Figure \ref{fig:beta}), which is centered close to the model distribution including differential extinction, but has a somewhat higher dispersion. This higher observed dispersion may be caused by uncertainties in both the modeled colors and the observed colors.  To investigate this possibility we add UV color shifts to the modeled FUV-NUV color which are drawn, for each galaxy, from a normal distribution with standard deviation given by the quadrature sum of the model and photometric uncertainties in UV color. Including these uncertainties, a two-sided Kolmogorov-Smirnoff test is unable to reject the hypothesis that the modeled and observed UV colors are drawn from the same distribution. Some additional small uncertainty in the modeled UV colors may arise from the differential reddening model assumed in the CMD analysis and our assumed extinction curve shape.

The blue peak of the UV color distribution in the top panel of Figure \ref{fig:beta} indicates that the SFH alone is not sufficient to explain the red offset in the UV colors of galaxies with low IRX, unless the recent SFHs are systematically incorrect. However, the middle and bottom panels of Figure \ref{fig:beta} indicate that the inclusion of differential extinction is sufficient to explain the observed UV color distribution well.

%%%%%%%%%%%%%%%%
%FIGURE - beta - SFH
%%%%%%%%%%%%%%%

\section{The Relationship Between UV luminosity and SFR}
\label{sec:model_sfr}
The UV luminosity of galaxies is commonly used as a tracer of the SFR both in the local universe and at high redshifts, where it is redshifted into the optical.  The conversion between UV luminosity and SFR is typically made assuming that the SFR is approximately constant for at least 100 Myr \citep{madau98,K98}.  However, if the SFR varies on shorter timescales, then the conversion between luminosity and SFR becomes more complicated, and the conversion depends on the exact distribution of stellar ages.  This complication has long been known \citep{K98}, but galaxies that undergo significant variations in their SFR on these timescales are thought to be rare.  There are two classes of galaxies where the assumption of a constant recent SFR is known to fail.  The first are starburst (and post-starburst) galaxies, which are thought to have enhanced recent SF due to  interactions \citep[e.g.,][]{hopkins06,wild09}.  The second are dwarf galaxies, for which rapid variation may be driven by stochasticity in the cluster formation process at low star formation rates \citep[e.g., ][]{fumagalli11}, or by the effects of feedback in low mass halos \citep[e.g., ][]{stinson07}. Sub-regions of galaxies also demonstrate variations in SFR on short timescales and episodic star formation \citep[e.g.,][]{gogarten09, huang13}.

The model luminosities determined in \S\ref{sec:model} allow us to explore the effect of realistic, observationally constrained SFHs on the ratio of UV luminosity to the SFR averaged over a given timescale.  We define the SFR conversion factor  $\epsilon_{\lambda,T}$ via the equation
\begin{displaymath}
\langle SFR\rangle_{T} = (\lambda \mbox{L}_\lambda)/\epsilon_{\lambda,T}
\end{displaymath}
 where $T$ is the log of the averaging timescale in years. The FUV luminosity is typically assumed to trace the SFR over 100 Myr, approximately the lifetime of FUV emitting stars.  We will thus consider $\epsilon_{FUV,8}$. The modeled FUV luminosity accounts for the effects of SFH, but not dust attenuation, metallicity, or other effects as discussed in \S\ref{sec:compare}. Because both the SFR and the modeled UV luminosity are derived from the same SFH, the results we obtain in this section do not involve the observed luminosity. Furthermore, our conclusions do not require that the derived SFHs (\S\ref{sec:data:sfhs}) are exactly correct in detail for each galaxy, only that they are plausible and not systematically biased.

%%%%%%%%%%%%%%%%
%FIGURE - epsilon^model distribution
%%%%%%%%%%%%%%%
\begin{figure}
\epsscale{1.2}
\plotone{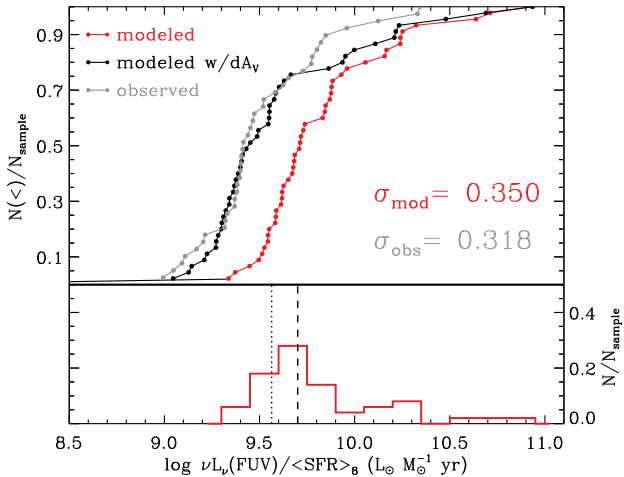}
\figcaption{The modeled FUV to SFR conversion factors for the sample. The distribution of $\epsilon_{FUV,8}^{mod}$, the ratio of the modeled intrinsic FUV luminosity to $\langle SFR \rangle_8$, is shown as a red histogram in the bottom panel, while the connected red points in the top panel show the cumulative distribution. The grey connected points show the cumulative distribution of $\epsilon_{FUV,8}^{obs}$, the ratio of the \emph{observed} FUV luminosity to \sfre, while the black connected points show the cumulative distribution of the ratio of the modeled luminosity including dust to \sfre.   The vertical dotted line marks the conversion factor of \cite{K98}, while the dashed line is the constant SFR conversion factor for the population synthesis models adopted here with 0.2Z$_\sun$ metallicity.  The expected conversion factors for exponentially declining, 0.2Z$_\sun$ metallicity models with $e$-folding times of 1 to 13 Gyr and ages more than 1 Gyr are all to the right of the dashed line. 
\label{fig:sfr_ratio_mod}}
\end{figure}

 In Figure \ref{fig:sfr_ratio_mod}, we present the distribution of $\epsilon_{FUV,8}^{mod}$, i.e. the ratio of the modeled FUV luminosity to \sfre.  We show this ratio for both the modeled intrinsic luminosities and for the modeled luminosities including differential reddening. The $\epsilon^{mod}_{FUV,8}$ that we derive can be compared directly to the conversion factor of \citet{K98}, plotted as the dashed line in Figure \ref{fig:sfr_ratio_mod}.  \cite{K98} derived a single conversion factor assuming solar metallicity, a nearly constant SFR, and the same IMF as we have adopted \citep[see also][]{madau98}. Figure \ref{fig:sfr_ratio_mod} shows that the observationally constrained SFH, on its own, induces significant scatter in $\epsilon_{FUV,8}$.  Depending on the SFH of a particular galaxy, the conversion between UV luminosity and \sfre~ can vary by an order of magnitude. For our sample of galaxies, the r.m.s. dispersion in $\epsilon^{mod}_{FUV,8}$ is a factor of $\sim 2$. 

The scatter in the ratio of \sfre~ to the modeled FUV luminosity is 0.3 dex, while the scatter in the ratio of the observed luminosity to the modeled FUV luminosity (including SFH effects) is only 0.14 dex.  This suggests that the SFH dominates the scatter in the conversion from FUV luminosity to \sfre~in this sample.  Additional effects such as photometric uncertainties and metallicity variations, and the distribution of dust attenuation, can only contribute an additional 0.14 dex of scatter.

%%%%%%%%%%%%%%%%
%FIGURE - epsilon^model versus color
%%%%%%%%%%%%%%%
\begin{figure*}
\plotone{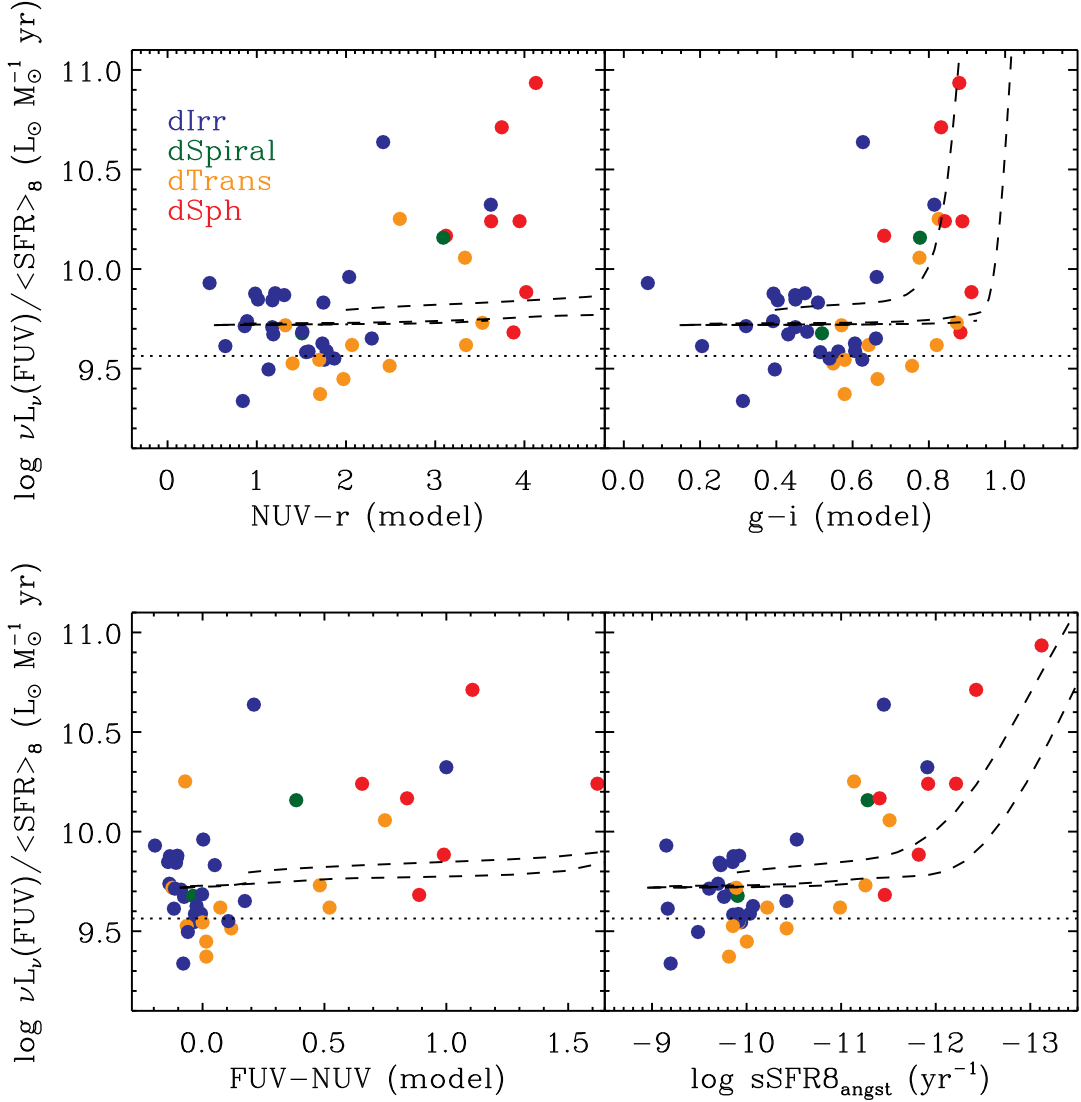}
\figcaption{Color dependence of the modeled FUV to SFR conversion factor. $Top left:$ The ratio of the modeled FUV luminosity to $\langle SFR \rangle_8$ (i.e. $\epsilon_{FUV,8}^{mod}$), as a function of the modeled $NUV-r$ color. Different morphological types are shown in different colors: dSph ($red$), dTrans ($orange$), dSpiral ($green$), and dI ($blue$). $Top right:$ The same, but plotted against the modeled $g-r$ color. $Bottom left:$ The same, but for the FUV$-$NUV color.  $Bottom right:$ The same, but plotted against the specific SFR derived from the SFH. The  dotted line in each panel marks the conversion factor of \cite{K98}, while the dashed lines are the expectations for population synthesis models with several different rates of exponential decline.
\label{fig:sfr_ratio_color}}
\end{figure*}

The conversion factor of \cite{K98} shown in Figure \ref{fig:sfr_ratio_mod} is lower than the median $\epsilon_{FUV,8}$ that we derive. This can be traced to the different assumptions made in \cite{K98} about the metallicity and the duration of star formation, and is not due to the details of the recent SFH. To better match the properties of our sample, we derive an alternative version of the \citep{K98} conversion factor as follows. We use the population synthesis models described in \S\ref{sec:model}, assuming $Z=0.2Z_\sun$ and constant SFR lasting $>10$ Gyr, and derive a value of the conversion factor that is larger than the \cite{K98} value, and that is approximately the median of $\epsilon_{FUV,8}^{mod}$. This larger value is due to the combined effects of lower metallicity and a longer period of constant SFR than adopted by \cite{K98}. 

In models with exponentially declining SFRs, the conversion factor $\epsilon_{FUV,8}$ varies with the specific SFR of the model.  In such $\tau$-models, $\epsilon_{FUV,8}$ increases as the specific SFR decreases, and is almost always larger than in the case of constant SFR\footnote{Conversion factors smaller than the constant SFR expectation are only possible in $\tau$-models when the age is very young, $\lesssim 300$ Myr}. Thus, if the SFHs were well described by smoothly evolving $\tau$-models, we would expect to find a correlation between the conversion factor and some tracer of the specific SFR, such as broadband color. In Figure \ref{fig:sfr_ratio_color} we show the color (and specific SFR) dependence of $\epsilon_{FUV,8}^{mod}$.  The much smoother relationships between $\epsilon_{FUV,8}$ and broadband colors that are expected for exponentially declining models are plotted as dashed lines in Figure \ref{fig:sfr_ratio_color}.  We find only a very weak correlation between $\epsilon_{FUV,8}^{mod}$ and broadband color.  That is, for the SFHs considered here, it is difficult to predict $\epsilon_{FUV,8}^{mod}$ from broadband color information alone, even when the effects of dust attenuation and metallicity variations are not included. The lack of correlation is due to the fact that $\epsilon_{FUV,8}$ traces a shorter timescale of SFH than the broadband colors, and that for our sample the SFH on short timescales is not well correlated with the SFH on longer timescales.

\subsection{The Timescale of UV emission}
\label{sec:cdf}

%%%%%%%%%%%%%%%%
%FIGURE - UV flux CDF
%%%%%%%%%%%%%%%
\begin{figure*}
\plotone{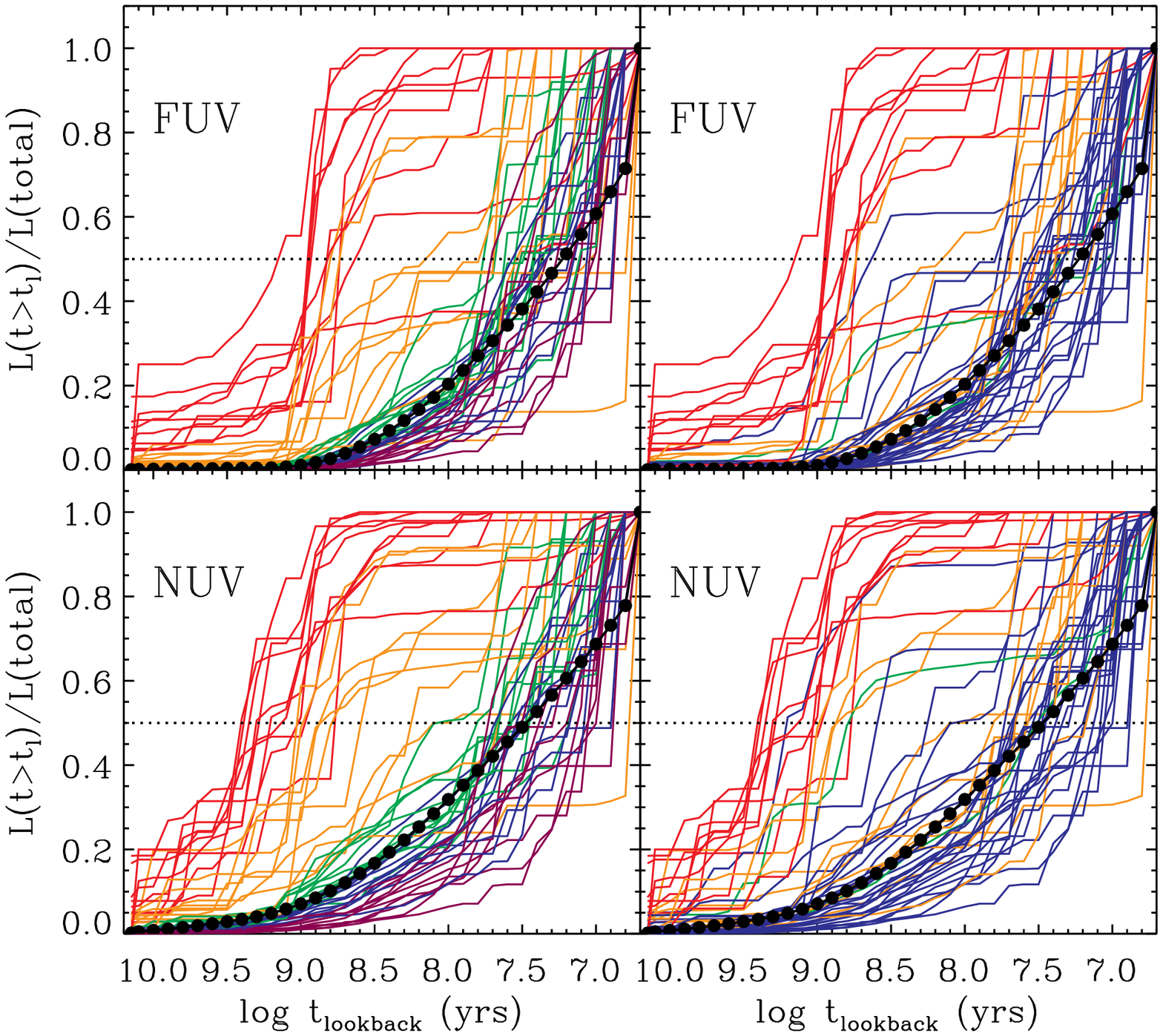}
\figcaption{The source of UV luminosity in galaxies. The colored lines give the cumulative fraction of current FUV ($top$) or NUV ($bottom$) luminosity produced by stars older than $t_{lookback}$. The black solid points are the expectation for constant SFR.  The dotted line denotes 50\% of the current UV luminosity.  $Left: $The color coding is by the modeled $NUV-r$ color, with bins in color of $0.05<NUV-r<1.1$ ($purple$) $1.1<NUV-r<1.48$ ($blue$) $1.48<NUV-r<1.86$ ($green$) $1.86<NUV-r<2.86$ ($orange$) $2.86<NUV-r<4.0$ ($red$).  $Right:$ the color coding is by morphological type, showing dIrr ($blue$), dSpiral ($green$), dTrans ($orange$) and dSph ($red$). 
\label{fig:cdf_flux_fuv}}
\end{figure*}

It is often assumed that the FUV and NUV luminosity trace the SFR over timescales of $\sim 100$ and $\sim 200$ Myr, respectively. However, this assumption is only valid for a constant SFR. We have demonstrated that the SFH of galaxies can cause significant variation in $\epsilon_{FUV,8}$, the ratio of the UV luminosity to the SFR averaged over 100 Myr. This is due to galaxy by galaxy variations in the distribution of the ages of stars contributing to the FUV luminosity.  A single representative timescale for the UV luminosity is therefore difficult to define for a sample of galaxies with diverse SFH.  It is important to understand this difficulty (and correct for it if possible) when comparing SFR determinations between samples of galaxies with different SFH.

With strong constraints on the SFH of galaxies from optical color-magnitude diagrams, it is possible to explore the age distribution of stars contributing to the UV luminosity of real galaxies, and to define a characteristic timescale for each galaxy.  In Figure \ref{fig:cdf_flux_fuv} we show, for each galaxy, the cumulative fraction of the current UV luminosity as a function of the age of the contributing stars (i.e., the fraction of the current UV luminosity that is produced by stars that are older than a given lookback time).  For a constant SFR this cumulative fraction shows an almost linear rise in the FUV, as nearly equal amounts of the current FUV luminosity are contributed by stars in equal logarithmic age bins, up to $\sim$100 Myr ages \citep[see also][]{K12}.  However, few of the sample galaxies display this behavior.  At the extremes, galaxies that reach a cumulative fraction of 1 at large lookback times have nearly all of their UV luminosity coming from relatively old stars.  Conversely, galaxies that reach a cumulative fraction of 1 at smaller lookback times have a larger contribution to the current UV luminosity from young stars.  

In Figure \ref{fig:cdf_flux_fuv} we see that the fraction of the total FUV luminosity that is produced by stars younger than 10 Myr ranges from 0\% to $\sim 60$\%.  The fraction of the total FUV luminosity that is produced by stars older than 100 Myr ranges from less than 5\% to 100\%.  Similarly, the fraction of the total NUV luminosity produced by stars older than 200 Myr ranges from less than 5\% to 100\%. 

The age distribution of UV emitting stars depends on the SFH, and so it may be possible to identify coarse measures of SFH that correlate with the cumulative fractions given in Figure \ref{fig:cdf_flux_fuv}.  We focus on the $NUV-r$ color, an observable that is often used as a proxy for the coarse, long-term SFH of a galaxy \citep{salim05, salim07, schiminovich07}. We have separated the galaxies in Figure \ref{fig:cdf_flux_fuv} by the predicted $NUV-r$ color. This separation shows that for galaxies with $NUV-r > 2.86$, more than 50\% of the modeled FUV  luminosity arises from stars older than 100 Myr.  For galaxies with $NUV-r < 1.1$, more than 50\% of the modled FUV luminosity is contributed by stars younger than $\sim 16$ Myr.  

We have also separated the galaxies by morphological type. Morphological type is correlated with $NUV-r$ color, and so we see a similar difference between galaxies of the dSph type (for which a majority of the FUV flux arises from stars older than 100Myr) and dIrr. The behavior in the NUV is similar, but shifted to larger timescales. 

%%%%%%%%%%%%%%%%
%FIGURE - Tau_50
%%%%%%%%%%%%%%%
\begin{figure*}
\plotone{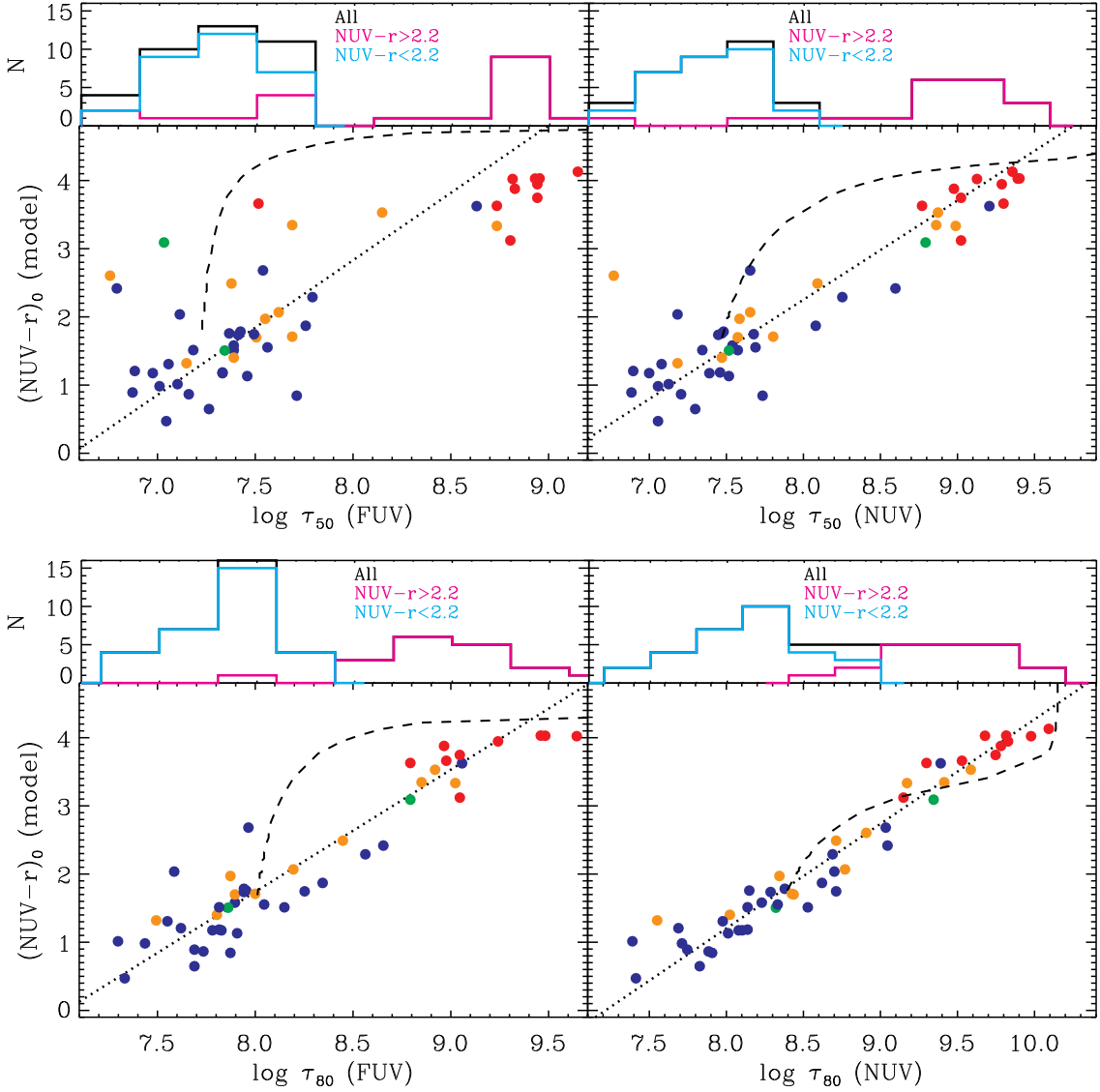}
\figcaption{The timescale probed by the UV luminosity.  \emph{Top left:} The modeled dust-free $(NUV-r)_0$ color is shown as a function of $\tau_{50,FUV}$.  Stars younger than $\tau_{50,FUV}$ contribute 50\% of the modeled present day FUV luminosity.  Points are color-coded by morphology: dSph ($red$), dTrans ($orange$), dSpiral ($green$), and dI ($blue$).  The dashed line gives $(NUV-r)_0$ color versus $\tau_{50,FUV}$ for exponentially declining models with different decline rates at ages of 14 Gyr. The dotted line shows the linear fit given in Table \ref{tbl:tau_fit}. The upper panel shows the distribution of $\tau_{50,FUV}$ for the entire sample ($black$) and for the sample split by modeled $(NUV-r)_0$ color ($magenta$ and $cyan$). \emph{Top right:} The same, but for $\tau_{50,NUV}$. {\it Bottom Left:}  As for the top right, but $(NUV-r)_0$ is plotted against $\tau_{80,FUV}$.  Stars younger than $\tau_{80,FUV}$ contribute 80\% of the modeled present day FUV flux.  {\it Bottom right:} The same but for $\tau_{80,NUV}$.
\label{fig:tau_50}}
\end{figure*}

We can use these cumulative luminosity fractions to construct, for each galaxy, a measure of the typical age of the UV producing population.  This measure, $\tau_{50}$, is defined as the age where 50\% of the luminosity in a given band comes from younger stars, and 50\% from older stars. In Figure \ref{fig:cdf_flux_fuv},  $\tau_{50}$ is the lookback time where a colored line crosses the thin dotted line marking 50\% of the total UV luminosity. Figure \ref{fig:cdf_flux_fuv} shows that $\tau_{50}$ ranges from $\sim 7$Myr to 1 Gyr for the FUV and from $\sim 7$Myr to 3 Gyr for the NUV.  The median $\tau_{50}$ for the FUV and NUV are 26 and 38 Myr respectively.

In the top left panel of Figure \ref{fig:tau_50} we show $\tau_{50}$ for the FUV band against the modeled $NUV-r$ color.  There is a good correlation; redder galaxies typically have larger $\tau_{50,FUV}$ in the FUV, though several galaxies show much lower $\tau_{50,FUV}$ for their color than would be expected from the general trend.  The top right panel of Figure \ref{fig:tau_50} shows $\tau_{50}$ for the NUV band.  Here the correlation is even stronger, though of course all the $\tau_{50,NUV}$ are shifted to larger lookback times than $\tau_{50,NUV}$.  In Table \ref{tbl:tau_fit} we present the results of linear fits to the relationships between $NUV-r$ color and both $\tau_{50}$ and $\tau_{80}$ for the \emph{GALEX} bands.

The observed relationship between $NUV-r$ and $\tau_{50}$ is different than expected for exponentially declining models, which are also shown in Figure \ref{fig:tau_50}.  The observed relation is more linear than for exponentially declining models, and the minimum $\tau_{50}$ is lower for some galaxies than the minimum of the exponentially declining models.  In the FUV, approximately 12 of the sample galaxies (25\% of the sample) have $\tau_{50,FUV}<16$ Myr, which is the value expected for constant SFR.  Furthermore, for 10 of the sample galaxies (20\%) $\tau_{50,FUV}$ is greater than 100 Myr, the canonical FUV timescale, and these galaxies have bluer $NUV-r$ colors than exponentially declining models with the same $\tau_{50,FUV}$.

\subsection{Summary}

In this section we have demonstrated that SFH causes significant scatter in $\epsilon_{FUV,8}$, the ratio of the FUV luminosity to \sfre. Comparison with the observed FUV luminosity indicates that variations of the SFH dominate the total scatter in $\epsilon_{FUV,8}$ for this sample of galaxies. The scatter in $\epsilon_{FUV,8}$  does not correlate strongly with UV/optical colors or specific SFR. 

We also calculated the age distribution of stars that contribute to the FUV and NUV luminosity. We find a broad range of characteristic ages $\tau_{50}$, where $\tau_{50}$ is defined such that stars younger than $\tau_{50}$ contribute 50\% of the total UV luminosity. For the FUV band, $\tau_{50}$ ranges from less than 16 Myr (for a quarter of our sample) to greater than 100 Myr (for one fifth of our sample).  This is a sharp contrast to the standard assumptions for a constant SFR.  We also found that, in contrast to $\epsilon_{FUV,8}$,  $\tau_{50}$ \emph{is} correlated with UV-optical color.  This difference arises because while very recent star formation episodes not only makes galaxies bluer in the UV-optical color and dominate the UV luminosity, they also effectively erase the signatures from past SFR.  Therefore, the SFR at ages older than $\sim\tau_{50}$ simply adds noise to $\epsilon_{FUV,8}$. In effect, $\tau_{50}$ is a rough estimate of the timescale over which the UV is measuring the SFR, and it correlates with NUV-r color. 

\section{Conclusions}
In this study we have combined SFHs derived from resolved stellar CMDs with population synthesis modeling to predict the UV through Near-IR broadband SED of $\sim 50$ dwarf galaxies drawn from the ANGST survey. We have compared these predicted SEDs to the observed SEDs on a galaxy-by-galaxy basis.  We have also used these predicted SEDs to determine the effect of realistic star formation histories on the conversion between UV luminosity and SFR, and to derive characteristic timescales for the UV emission.  

{\it Summary of comparison to observations:} The comparison of the predicted SEDs to the observed SEDs reveals excellent agreement in the optical portion of the spectrum ($U$ through $i$ bands).  This agreement lends support to the accuracy of the derived SFHs, though it is in some part expected given that the resolved stars used to derive the SFH typically contribute $\sim40$\% of the total flux of the optical bands. In the \emph{GALEX} UV bands we also find very good agreement between the predicted and observed luminosities when including a model for differential attenuation that does not include a significant scattering component.

Using \emph{Spitzer} Near-IR photometry we have extended the comparison of population synthesis models to data to longer wavelength than in \cite{melbourne12}.  We find significant discrepancies between the predicted and observed integrated fluxes when using the \cite{marigo08} isochrones for TP-AGB stars.  However, we are unable to distinguish the different stellar contributors at these wavelengths, limiting our conclusions as to the source of discrepancy. Nevertheless, we find that modifications to the TP-AGB isochrones as suggested in \cite{conroy10_III} result in much better agreement between the observed and predicted luminosities at 3.6 and 4.5\micron.

We found that the differential extinction model that provides good matches to the optical CMD implies a larger than observed IR luminosity, under the assumption that all extinguished light is re-emitted in the IR.  Including the effects of scattering can reduce this discrepancy, but at the cost of introducing a discrepancy in the FUV and NUV bands.  We used the SFHs to calculate dust free FUV-NUV colors and compared these both to UV colors predicted when including differential attenuation and to the observed UV colors, finding that variations in the derived SFHs of these galaxies are not sufficient to explain the average observed UV color and its scatter.  The observed UV colors are consistent with differential extinction by dust. 

{\it Summary of model based results:}  Having gained confidence in the plausibility of the SFHs through comparison with observations -- while keeping in mind the discrepancies -- we have used the SFHs to determine the effect of realistic star formation histories on the conversion between UV luminosity and SFR.  We found that differences in the star formation histories of these dwarf galaxies could cause variations in the conversion between FUV luminosity and the \sfre~(the SFR averaged over 100 Myr) of an order of magnitude, with factor of two dispersion among individual galaxies in our sample.  For our sample of dwarf galaxies SFH is likely the dominant cause of scatter in the ratio of observed UV luminosity to the SFR averaged over 100 Myr, above variations in dust attenuation or metallicity.  We found that the variations in conversion from FUV luminosity to \sfre~were only poorly correlated with broadband color, due to the lack of smoothness in the SFH.

Using the SFHs of the individual galaxies we found that for significant numbers of our sample galaxies 50\% of the FUV luminosity is produced by stars younger than 16 Myr, while for 10 of the sample galaxies 50\% of the FUV luminosity was produced by stars older than 100 Myr.  This timescale at which 50\% of the luminosity is produced correlates with UV-optical color.  We found also that the fraction of FUV luminosity produced by stars older than 100 Myr ranges from less than 5\% to 100\%.

{\it Implications for  SFR measurement: }  There is a factor of two dispersion in the ratio of modeled UV luminosity to \sfre~that is due to SFH variations on short timescales.   This large dispersion is in contrast to common conversions from FUV luminosity to SFR that adopt a constant SFH and predict zero dispersion. The dispersion that we find is not correlated with galaxy color, which suggests that it will be difficult to improve on the estimate of \sfre~from UV luminosity or full SED modeling unless the SFH on short timescales can be otherwise constrained.  \cite{lee10} have shown, using SFHs drawn from semi-analytic models, that modeling the SED of high-redshift galaxies with an assumed SFH that does not match the true SFH can lead to biases and scatter in derived parameters.  We have extended this result to low-redshift dwarf galaxies. Our factor of two dispersion is similar to the scatter in recovered versus input SFR that \cite{lee10} derive when using UV luminosity as a monochromatic SFR indicator (note that the biases and scatter in the recovered SFR of these simulated high-redshift galaxies do not improve when fitting the entire SED). 

Many current SED fitting procedures utilize large grids of models that include bursts of star-formation superimposed on a more smooth evolution \citep[e.g,][]{brinchmann04, salim07, walcher08, moustakas11, pacifici12}.  However, unless the age, duration, and amplitude of the bursts can be well constrained -- which is exceedingly difficult even with optical spectroscopy -- it will be impossible to choose the correct model and hence derive the correct SFR for an individual galaxy.  It may be possible to properly incorporate the uncertainty on the derived SFR that is induced by short-term SFH variations.  However, even properly accounting for the uncertainty in the derived SFR (much less obtaining the correct SFR for individual galaxies) requires that the `burst' parameters (frequency, amplitude distribution, and durations) used in the library are representative of the population being fit \citep{sedfit}, whereas in practice they are rather arbitrarily defined \citep[though see][for a case where the library itself is based on semi-analytics models]{pacifici12}. The strong constraints on the SFH provided by color-magnitude diagrams of resolved stars have allowed us to accurately assess the uncertainty in derived SFR for a sample of nearby dwarf galaxies.

{\it Applicability to other populations:}  It is important to consider the applicability of these results to different galaxy populations.  The sample of galaxies we have considered is characterized by low stellar masses, low metallicities, low SFRs, and more stochastic SF. In more massive star forming galaxies, either locally or at high redshift, the recent SFH is likely to be more smooth than in the dwarfs of the present study.  The scatter in the conversion between UV luminosity and SFR would then be smaller than we observe.  However, when considering small parts of larger galaxies, for example to generate SFR maps, such SFH variations may be present and induce significant uncertainty in maps of SFR derived from UV luminosities  \citep[e.g.,][]{bigiel08,leroy12}.

\acknowledgements

Based on observations made with the NASA/ESA Hubble Space Telescope, obtained at the Space Telescope Science Institute, which is operated by the Association of Universities for Research in Astronomy, Inc., under NASA contract NAS 5-26555. These observations are associated with programs \#9771, 9884, 10210, 10503, 10605, 10915, and 11986.  We gratefully acknowledge NASA's support for construction, operation, and science analysis for the GALEX mission, developed in cooperation with the Centre National d’Etudes Spatiales of France and the Korean Ministry of Science and Technology. This work is based in part on observations made with the Spitzer Space Telescope, which is operated by the Jet Propulsion Laboratory, California Institute of Technology under a contract with NASA. This research has made extensive use of NASA's Astrophysics Data System Bibliographic Services. This research has made use of the NASA/IPAC Extragalactic Database, which is operated by JPL/Caltech, under contract with NASA.

Funding for SDSS-III has been provided by the Alfred P. Sloan Foundation, the Participating Institutions, the National Science Foundation, and the U.S. Department of Energy Office of Science. The SDSS-III web site is http://www.sdss3.org/.

SDSS-III is managed by the Astrophysical Research Consortium for the Participating Institutions of the SDSS-III Collaboration including the University of Arizona, the Brazilian Participation Group, Brookhaven National Laboratory, University of Cambridge, Carnegie Mellon University, University of Florida, the French Participation Group, the German Participation Group, Harvard University, the Instituto de Astrofisica de Canarias, the Michigan State/Notre Dame/JINA Participation Group, Johns Hopkins University, Lawrence Berkeley National Laboratory, Max Planck Institute for Astrophysics, Max Planck Institute for Extraterrestrial Physics, New Mexico State University, New York University, Ohio State University, Pennsylvania State University, University of Portsmouth, Princeton University, the Spanish Participation Group, University of Tokyo, University of Utah, Vanderbilt University, University of Virginia, University of Washington, and Yale University.

%%%%%%%%%%%%%%%%%%%%

\bibliographystyle{apj}

\clearpage

\begin{deluxetable}{lcccccccc}
\tablecolumns{9}
\tablecaption{Galaxy sample properties \label{tbl:pars}}
\tablehead{\colhead{Name} & \colhead{Alternate}& \colhead{$\log\langle \mbox{SFR}\rangle_8$\tablenotemark{a}} & \colhead{$\log\mbox{M}_*$\tablenotemark{a}} & \colhead{$A_{fuv}$\tablenotemark{b}} & \colhead{E$(B-V)_{MW}$} & \colhead{$\log[\mbox{O}/\mbox{H}]+12$\tablenotemark{c}} & \colhead{f$_{HST}$\tablenotemark{d}} & \colhead{T type}  \\
\colhead{} & \colhead{Name} & \colhead{(M$_\sun$ yr$^{-1}$)} & \colhead{(M$_\sun$)} & \colhead{(mag)} & \colhead{} & \colhead{(B12)} & \colhead{} & \colhead{} }
\startdata
  AM1001-270    &          Antlia &   -4.086$^{+ 0.400}_{- 0.116}$ &    6.666$^{+ 0.025}_{- 0.021}$ & $<   0.34$ &    0.079 &   7.41 &   1.00 & 10 \\
  BK5N          &         \nodata &   -4.706$^{+ 0.372}_{- 0.203}$ &    7.293$^{+ 0.009}_{- 0.024}$ & $<   4.78$ &    0.063 &   7.55 &   1.00 & -3 \\
  UGCA276       &          DDO113 &   -4.788$^{+ 0.216}_{- 0.270}$ &    7.129$^{+ 0.008}_{- 0.013}$ & $<   0.12$ &    0.020 &   7.57 &   1.00 & 10 \\
  UGC7577       &          DDO125 &   -2.171$^{+ 0.028}_{- 0.019}$ &    7.775$^{+ 0.016}_{- 0.013}$ &    0.04$\pm$  0.006 &    0.020 &   7.78 &   0.77 & 10 \\
  UGC8651       &          DDO181 &   -2.456$^{+ 0.018}_{- 0.018}$ &    7.580$^{+ 0.003}_{- 0.010}$ &    0.04$\pm$  0.006 &    0.006 &   7.67 &   0.97 & 10 \\
  UGC8760       &          DDO183 &   -2.349$^{+ 0.016}_{- 0.014}$ &    7.567$^{+ 0.011}_{- 0.011}$ &    0.02$\pm$  0.003 &    0.016 &   7.70 &   1.00 & 10 \\
  UGC9128       &          DDO187 &   -2.880$^{+ 0.034}_{- 0.023}$ &    7.039$^{+ 0.006}_{- 0.007}$ & $<   0.11$ &    0.023 &   7.56 &   1.00 & 10 \\
  UGC9240       &          DDO190 &   -2.102$^{+ 0.013}_{- 0.009}$ &    7.791$^{+ 0.005}_{- 0.011}$ &    0.14$\pm$  0.015 &    0.012 &   7.77 &   0.96 & 10 \\
  UGCA133       &          DDO044 &   -4.795$^{+ 0.148}_{- 0.505}$ &    7.420$^{+ 0.009}_{- 0.004}$ & $<   4.92$ &    0.041 &   7.65 &   1.00 & -3 \\
  UGCA015       &          DDO006 &   -2.460$^{+ 0.007}_{- 0.014}$ &    7.351$^{+ 0.009}_{- 0.005}$ & $<   0.17$ &    0.017 &   7.60 &   1.00 & 10 \\
  DDO078        &         \nodata &   -4.676$^{+ 0.191}_{- 0.316}$ &    7.782$^{+ 0.009}_{- 0.012}$ & $<   3.62$ &    0.021 &   7.49 &   0.41 & -3 \\
  UGC5692       &          DDO082 &   -2.876$^{+ 0.039}_{- 0.037}$ &    8.404$^{+ 0.009}_{- 0.003}$ &    0.26$\pm$  0.032 &    0.041 &   7.90 &   0.67 &  9 \\
  UGC6817       &          DDO099 &   -2.288$^{+ 0.018}_{- 0.030}$ &    7.565$^{+ 0.010}_{- 0.059}$ &    0.05$\pm$  0.007 &    0.026 &   7.67 &   0.74 & 10 \\
  ESO294-G010   &         \nodata &   -4.118$^{+ 0.081}_{- 0.126}$ &    7.024$^{+ 0.003}_{- 0.005}$ & $<   0.55$ &    0.006 &   7.56 &   0.96 & -3 \\
  ESO410-G005   &         \nodata &   -3.901$^{+ 0.089}_{- 0.057}$ &    7.106$^{+ 0.005}_{- 0.004}$ & $<   0.55$ &    0.014 &   7.59 &   1.00 & -1 \\
  ESO540-G032   &         \nodata &   -3.864$^{+ 0.091}_{- 0.095}$ &    7.415$^{+ 0.009}_{- 0.006}$ & $<   1.14$ &    0.020 &   7.52 &   1.00 & -3 \\
  F08D1         &         \nodata &   -4.573                       &    8.007$^{+ 0.012}_{- 0.003}$ & $<   0.92$ &    0.108 &   7.80 &   1.00 & -3 \\
  UGC8091       &             GR8 &   -2.887$^{+ 0.032}_{- 0.030}$ &    6.966$^{+ 0.020}_{- 0.024}$ &    0.04$\pm$  0.005 &    0.026 &   7.54 &   1.00 & 10 \\
  HS117         &        HS98-117 &   -5.017$^{+ 0.332}_{- 0.201}$ &    6.548$^{+ 0.005}_{- 0.014}$ & $<   2.65$ &    0.115 &   7.50 &   0.97 & 10 \\
  UGC5666       &          IC2574 &   -1.084$^{+ 0.003}_{- 0.003}$ &    8.821$^{+ 0.002}_{- 0.002}$ &    0.14$\pm$  0.016 &    0.036 &   8.09 &   0.74 &  9 \\
  IKN           &         \nodata &   -3.643$^{+ 0.049}_{- 0.111}$ &    7.812$^{+ 0.011}_{- 0.016}$ & $<   4.67$ &    0.058 &   7.51 &   1.00 & -3 \\
  M81-DwA       &           KDG52 &   -2.715$^{+ 0.021}_{- 0.014}$ &    7.146$^{+ 0.005}_{- 0.008}$ &    0.20$\pm$  0.029 &    0.020 &   7.48 &   1.00 & 10 \\
  KDG61         &         KK98-81 &   -3.995$^{+ 0.102}_{- 0.100}$ &    7.675$^{+ 0.010}_{- 0.012}$ & $<   3.28$ &    0.073 &   7.69 &   0.99 & -1 \\
  UGC5442      &            KDG64 &   -5.367                       &    7.563$^{+ 0.005}_{- 0.008}$ &    1.77$\pm$  1.717 &    0.053 &   7.71 &   1.00 & -3 \\
  KDG73         &         \nodata &   -3.071$^{+ 0.025}_{- 0.028}$ &    6.923$^{+ 0.025}_{- 0.012}$ & $<   0.83$ &    0.019 &   7.52 &   1.00 & 10 \\
  KKR03         &        KK98-230 &   -3.697$^{+ 0.064}_{- 0.060}$ &    6.222$^{+ 0.014}_{- 0.011}$ & $<   0.37$ &    0.014 &   7.31 &   1.00 & 10 \\
  KKH037        &        HS98-010 &   -3.255$^{+ 0.042}_{- 0.040}$ &    7.376$^{+ 0.010}_{- 0.012}$ & $<   0.49$ &    0.074 &   7.60 &   1.00 & 10 \\
  KKH086        &         \nodata &   -3.927$^{+ 0.160}_{- 0.099}$ &    6.525$^{+ 0.038}_{- 0.023}$ & $<   1.05$ &    0.027 &   7.45 &   1.00 & 10 \\
  KKH098        &         \nodata &   -3.298$^{+ 0.052}_{- 0.026}$ &    6.930$^{+ 0.018}_{- 0.027}$ & $<   0.22$ &    0.123 &   7.52 &   1.00 & 10 \\
  KKR25         &         \nodata &   -5.267$^{+ 0.352}_{- 1.142}$ &    6.322$^{+ 0.025}_{- 0.033}$ & $<   4.54$ &    0.009 &   7.29 &   0.89 & 10 \\
  NGC2366       &         UGC3851 &   -1.176$^{+ 0.003}_{- 0.004}$ &    8.585$^{+ 0.002}_{- 0.002}$ &    0.40$\pm$  0.039 &    0.036 &   7.96 &   0.87 & 10 \\
  NGC3741       &         UGC6572 &   -2.349$^{+ 0.019}_{- 0.014}$ &    7.517$^{+ 0.006}_{- 0.009}$ &    0.05$\pm$  0.006 &    0.024 &   7.67 &   0.98 & 10 \\
  NGC4163       &         UGC7199 &   -2.615$^{+ 0.022}_{- 0.015}$ &    7.927$^{+ 0.002}_{- 0.007}$ &    0.12$\pm$  0.014 &    0.020 &   7.79 &   1.00 & 10 \\
  UGC4483       &         \nodata &   -2.490$^{+ 0.021}_{- 0.027}$ &    7.214$^{+ 0.007}_{- 0.017}$ &    0.10$\pm$  0.011 &    0.034 &   7.53 &   1.00 & 10 \\
  UGC8201       &          DDO165 &   -1.220$^{+ 0.004}_{- 0.005}$ &    8.267$^{+ 0.011}_{- 0.030}$ &    0.02$\pm$  0.004 &    0.024 &   7.86 &   1.00 & 10 \\
  UGC8508       &           IZw60 &   -2.546$^{+ 0.016}_{- 0.015}$ &    7.377$^{+ 0.003}_{- 0.004}$ & $<   0.00$ &    0.015 &   7.67 &   0.94 & 10 \\
  UGC8833       &         \nodata &   -2.779$^{+ 0.024}_{- 0.025}$ &    7.294$^{+ 0.010}_{- 0.016}$ & $<   0.16$ &    0.012 &   7.61 &   1.00 & 10 \\
  UGCA292       &         \nodata &   -2.486$^{+ 0.030}_{- 0.016}$ &    6.688$^{+ 0.022}_{- 0.031}$ &    0.06$\pm$  0.010 &    0.016 &   7.49 &   1.00 & 10 \\
  UGCA438       &     ESO407-G018 &   -2.582$^{+ 0.030}_{- 0.020}$ &    7.306$^{+ 0.052}_{- 0.011}$ & $<   0.07$ &    0.014 &   7.66 &   1.00 & 10 \\
  UGC5428       &          DDO071 &   -5.125                       &    7.535$^{+ 0.003}_{- 0.007}$ & $<   0.00$ &    0.098 &   7.66 &   0.98 & -3 \\
  Sex A         &          DDO075 &   -2.370$^{+ 0.009}_{- 0.010}$ &    6.793$^{+ 0.006}_{- 0.015}$ &    0.03$\pm$  0.003 &    0.045 &   7.70 &   0.31 & 10 \\
  ESO321-G014   &         \nodata &   -2.800$^{+ 0.047}_{- 0.034}$ &    7.286$^{+ 0.011}_{- 0.016}$ &    0.01$\pm$  0.011 &    0.094 &   7.67 &   0.96 & 10 \\
  ESO540-G030   &           KDG02 &   -4.764$^{+ 0.327}_{- 0.305}$ &    6.690$^{+ 0.010}_{- 0.020}$ & $<   1.28$ &    0.023 &   7.60 &   1.00 & -1 \\
  IC5152        &     ESO237-G027 &   -1.962$^{+ 0.009}_{- 0.013}$ &    7.746$^{+ 0.015}_{- 0.007}$ &    0.14$\pm$  0.015 &    0.025 &   7.95 &   0.39 & 10 \\
  Sculptor-dE1  &            Sc22 &   -4.642$^{+ 0.167}_{- 0.326}$ &    7.168$^{+ 0.012}_{- 0.014}$ & $<   5.07$ &    0.015 &   7.50 &   0.99 & -3 \\
  UGC5373       &   Sex B, DDO070 &   -2.753$^{+ 0.023}_{- 0.020}$ &    7.126$^{+ 0.006}_{- 0.013}$ &    0.05$\pm$  0.006 &    0.031 &   7.74 &   0.33 & 10 \\
  UGC4305       &   Ho II, DDO050 &   -1.173$^{+ 0.003}_{- 0.002}$ &    8.432$^{+ 0.001}_{- 0.002}$ &    0.11$\pm$  0.012 &    0.032 &   8.00 &   0.80 & 10 \\
  UGC4459       &          DDO053 &   -2.299$^{+ 0.017}_{- 0.015}$ &    7.687$^{+ 0.010}_{- 0.012}$ &    0.21$\pm$  0.023 &    0.038 &   7.69 &   1.00 & 10 \\
  UGC5139       &    Ho I, DDO063 &   -1.799$^{+ 0.006}_{- 0.010}$ &    8.059$^{+ 0.005}_{- 0.006}$ &    0.04$\pm$  0.006 &    0.050 &   7.78 &   0.97 & 10 \\
  UGC5336       &   Ho IX, DDO066 &   -1.529$^{+ 0.007}_{- 0.005}$ &    7.668$^{+ 0.010}_{- 0.012}$ &    0.03$\pm$  0.005 &    0.079 &   7.71 &   0.98 & 10 \\
\enddata
\tablenotetext{a}{Tabulated central values are derived from the best-fit SFH. Upper and lower error bars correspond to the 84th and 16th percentiles, respectively, of the distribution of values derived from the Monte-Carlo realizations of the SFH. The best-fit \sfre~ is occasionally zero; in these cases we report the 84th percentile of the distribution of Monte-Carlo values (an approximate 1$\sigma$ upper limit) and no error bars are given.}
\tablenotetext{b}{Internal dust attenuation estimated from the ratio of the 24\micron\ to FUV flux following \citet{hao11}}
\tablenotetext{c}{Gas-phase metallicity as estimated from the 4.5\micron\ luminosity-metallicity relation of \citet{berg12}}
\tablenotetext{d}{Fraction of the $g$ or $B$ band flux within the \citet{dale_lvl} apertures that also falls within the HST FOV, i.e., an aperture correction.}
\end{deluxetable}

\begin{deluxetable}{lcccccccccccc}
\tablecolumns{13}
\tablecaption{Integrated Photometry Within the HST Footprint\label{tbl:photom}}
\tablehead{\colhead{Name} & ($m-M$)\tablenotemark{a} & \colhead{$FUV$}& \colhead{$NUV$}& \colhead{$u (U)$\tablenotemark{b}}& \colhead{$g (B)$ \tablenotemark{b}}& \colhead{$r (V)$ \tablenotemark{b}}& \colhead{$i (R)$ \tablenotemark{b}}& \colhead{$z$}& \colhead{$3.6\micron$}& \colhead{$4.5\micron$}& \colhead{$5.6\micron$}& \colhead{$8\micron$}\\
\colhead{} & \colhead{mag} & \colhead{mag} & \colhead{mag} & \colhead{mag} & \colhead{mag} & \colhead{mag} & \colhead{mag} & \colhead{mag} & \colhead{mag} & \colhead{mag} & \colhead{mag} & \colhead{mag} }
\startdata
  AM1001-270    &  25.40 &   -7.79 &   -8.79 & (  -8.26) & ( -10.26) & ( -10.71) & ( -10.77) & \nodata &  -10.38 &   -9.93 &  \nodata &   -9.67 \\
  BK5N         &  27.89 &   -7.89 &   -7.56 &  \nodata &  \nodata &  \nodata &  \nodata &  \nodata &  -11.98 &  -11.09 &  \nodata &  -12.56 \\
  UGCA276       &  27.35 &   -7.08 &   -8.92 &  \nodata &  \nodata &  \nodata &  \nodata &  \nodata &  -12.12 &  -11.52 &  -10.51 &  \nodata \\
  UGC7577       &  27.02 &  -12.01 &  -12.35 &  -13.31 &  -14.12 &  -14.46 &  -14.62 &  -14.70 &  -13.78 &  -13.19 &  -12.58 &  -12.17 \\
  UGC8651       &  27.40 &  -11.62 &  -11.83 &  -12.55 &  -13.32 &  -13.58 &  -13.71 &  -13.87 &  -13.02 &  -12.52 &  -10.87 &  -11.14 \\
  UGC8760       &  27.51 &  -11.57 &  -11.90 &  -12.58 &  -13.38 &  -13.59 &  -13.82 &  -13.74 &  -13.13 &  -12.75 &  -12.31 &  \nodata \\
  UGC9128       &  26.72 &  -10.43 &  -10.82 &  -11.79 &  -12.58 &  -12.78 &  -12.90 &  -12.95 &  -12.09 &  -11.78 &  -10.20 &  -11.36 \\
  UGC9240       &  27.23 &  -12.30 &  -12.63 &  -13.34 &  -14.23 &  -14.49 &  -14.64 &  -14.85 &  -13.81 &  -13.39 &  -13.11 &  -13.19 \\
  UGCA133       &  27.45 &   -6.76 &   -9.43 &  -11.75 &  -12.65 &  -13.11 &  -13.16 &  -13.50 &  -12.92 &  -12.59 &  -10.17 &  \nodata \\
  UGCA015        &  27.60 &  -10.75 &  -10.96 & ( -11.78) & ( -12.38) & ( -12.57) & \nodata & \nodata &  -12.24 &  -11.94 &  -10.98 &  -10.50 \\
  DDO078        &  27.82 &  \nodata &  \nodata & \nodata & \nodata & \nodata & \nodata & \nodata &  \nodata &  \nodata &  \nodata &  \nodata \\
  UGC5692       &  28.06 &  -11.45 &  -12.36 & ( -14.20) & ( -14.91) & ( -15.45) & ( -16.55) & \nodata &  -15.40 &  -14.97 &  -15.07 &  -14.59 \\
  UGC6817       &  27.11 &  -11.82 &  -12.08 &  -12.68 &  -13.50 &  -13.71 &  -13.84 &  -13.76 &  -12.89 &  -12.31 &  -11.92 &  -10.67 \\
  ESO294-G010   &  26.47 &   -8.41 &   -8.86 & \nodata & ( -11.14) & \nodata & ( -11.99) & \nodata &  -11.85 &  -11.37 &  -10.50 &  \nodata \\
  ESO410-G005   &  26.42 &   -7.93 &   -9.23 & \nodata & ( -11.53) & \nodata & ( -12.25) & \nodata &  -12.97 &  -11.97 &  -12.49 &  -12.09 \\
  ESO540-G032   &  27.67 &   -8.14 &   -9.11 & \nodata & ( -11.40) & \nodata & ( -12.26) & \nodata &  -11.79 &  -10.97 &  \nodata &  \nodata \\
  F08D1          &  27.78 &  \nodata &   -8.94 &  \nodata &  \nodata &  \nodata &  \nodata &  \nodata &  -13.12 &  -12.82 &  -13.77 &  -13.26 \\
  UGC8091       &  26.60 &  -11.36 &  -11.36 &  -11.65 &  -12.15 &  -12.38 &  -12.43 &  -12.66 &  -11.53 &  -11.21 &  -10.79 &  -10.82 \\
  HS117      &  26.61 &   -7.74 &  \nodata & (  -8.98) & ( -10.69) & ( -11.49) & ( -12.02) & \nodata &  -11.59 &  -10.97 &  \nodata &  \nodata \\
  UGC5666       &  27.93 &  -15.38 &  -15.51 & ( -16.32) & ( -16.77) & ( -16.98) & ( -17.10) & \nodata &  -16.68 &  -16.31 &  -15.94 &  -16.01 \\
  IKN           &  27.84 &  \nodata &  \nodata & \nodata & \nodata & \nodata & \nodata & \nodata &  \nodata &  \nodata &  \nodata &  \nodata \\
  M81-DwA        &  27.68 &  -10.28 &  -10.42 &  \nodata &  \nodata &  \nodata &  \nodata &  \nodata &  -11.85 &  -11.00 &   -7.64 &  -10.36 \\
  KDG61         &  27.84 &   -8.81 &   -9.67 &  -12.78 &  -13.20 &  -13.83 &  -14.10 &  -14.38 &  -13.43 &  -12.85 &  \nodata &  \nodata \\
  UGC5442      &  27.84 &  \nodata &   -9.70 &  \nodata &  \nodata &  \nodata &  \nodata &  \nodata &  -13.20 &  -12.77 &  -10.38 &  -12.43 \\
  KDG73         &  27.84 &   -9.08 &   -9.35 &  \nodata &  \nodata &  \nodata &  \nodata &  \nodata &  -12.26 &  -10.95 &  \nodata &  -10.99 \\
  KKR03         &  26.41 &   -7.94 &   -8.42 &   -9.14 &   -9.94 &  -10.23 &  -10.18 &  -10.61 &  \nodata &  \nodata &  \nodata &  \nodata \\
  KKH037         &  27.65 &   -9.30 &   -9.82 & ( -11.21) & ( -12.09) & ( -12.51) & ( -12.81) & \nodata &  -12.33 &  -11.84 &  -11.47 &  -11.39 \\
  KKH086        &  27.03 &   -7.71 &   -8.76 &   -9.96 &  -10.74 &  -11.26 &  -11.33 &  -10.91 &  \nodata &  \nodata &  \nodata &  \nodata \\
  KKH098        &  26.90 &   -9.35 &  -10.08 &  -11.00 &  -11.81 &  -12.02 &  -12.39 &  -12.31 &  \nodata &  \nodata &  \nodata &  \nodata \\
  KKR25         &  26.35 &  \nodata &  \nodata &  \nodata &  \nodata &  \nodata &  \nodata &  \nodata &  \nodata &  \nodata &  \nodata &  \nodata \\
  NGC2366       &  27.53 &  -14.87 &  -15.03 & ( -15.77) & ( -16.06) & ( -16.26) & ( -16.30) & \nodata &  -15.71 &  -15.34 &  -15.52 &  -15.71 \\
  NGC3741       &  27.55 &  -12.32 &  -12.50 &  -12.95 &  -13.51 &  -13.73 &  -13.83 &  -14.02 &  -12.95 &  -12.51 &  -11.78 &  -11.44 \\
  NGC4163       &  27.36 &  -11.93 &  -12.37 &  -13.37 &  -14.22 &  -14.61 &  -14.86 &  -14.94 &  -14.09 &  -13.62 &  -13.23 &  -12.78 \\
  UGC4483       &  27.53 &  -11.71 &  -11.78 & ( -12.51) & ( -12.82) & ( -12.82) & ( -12.90) & \nodata &  -11.92 &  -11.14 &   -9.28 &  -11.09 \\
  UGC8201       &  28.35 &  -13.81 &  -14.03 &  -14.56 &  -15.29 &  -15.47 &  -15.55 &  -15.51 &  -14.64 &  -14.22 &  -13.79 &  -13.02 \\
  UGC8508       &  27.06 &  \nodata &  \nodata &  -12.63 &  -13.32 &  -13.63 &  -13.79 &  -13.94 &  -13.07 &  -12.55 &  -12.45 &  -12.42 \\
  UGC8833       &  27.47 &  -10.85 &  -11.05 &  -11.80 &  -12.55 &  -12.86 &  -13.02 &  -12.64 &  -12.33 &  -12.17 &  -11.47 &  -11.22 \\
  UGCA292       &  27.79 &  -11.44 &  -11.49 &  -12.13 &  -12.20 &  -12.38 &  -12.60 &  -12.61 &  -12.53 &  -11.59 &  -11.48 &  -13.03 \\
  UGCA438       &  26.74 &  -11.23 &  -11.37 & ( -11.69) & ( -12.46) & ( -13.00) & ( -13.15) & \nodata &  -12.65 &  -12.17 &  -11.96 &  -11.32 \\
  UGC5428       &  27.72 &  \nodata &   -9.87 & ( -11.34) & ( -12.52) & ( -13.02) & ( -13.30) & \nodata &  -12.94 &  -12.44 &  -11.26 &  -12.79 \\
  Sex A      &  25.60 &  -11.75 &  -11.87 & ( -12.27) & ( -12.53) & ( -12.56) & ( -12.66) & \nodata &  -11.91 &  -11.45 &  -10.90 &   -9.24 \\
  ESO321-G014   &  27.50 &  -10.93 &  -11.31 & ( -12.20) & ( -12.75) & ( -13.04) & ( -13.21) & \nodata &  -12.89 &  -12.48 &  -10.48 &  -11.56 \\
  ESO540-G030   &  27.61 &   -8.20 &   -9.31 & \nodata & ( -11.34) & \nodata & ( -10.38) & \nodata &  -12.01 &  -11.68 &  -10.13 &  \nodata \\
  IC5152        &  26.58 &  -13.35 &  -13.69 & ( -14.41) & ( -14.95) & ( -15.27) & ( -15.42) & \nodata &  \nodata &  -14.53 &  \nodata &  -15.31 \\
  Sculptor-dE1  &  28.07 &  \nodata &  \nodata & \nodata & \nodata & \nodata & \nodata & \nodata &  \nodata &  \nodata &  \nodata &  \nodata \\
  UGC5373       &  25.67 &  -10.76 &  -10.99 &  -11.72 &  -12.74 &  -13.03 &  -13.14 &  -13.25 &  -12.31 &  -11.93 &  \nodata &  -11.26 \\
  UGC4305       &  27.65 &  -15.00 &  -15.11 & ( -15.78) & ( -16.10) & ( -16.18) & ( -16.41) & \nodata &  -15.72 &  -15.45 &  -15.37 &  -15.45 \\
  UGC4459       &  27.79 &  -12.41 &  -12.57 &  -13.16 &  -13.68 &  -13.95 &  -14.06 &  -14.04 &  -13.56 &  -13.14 &  -12.80 &  -13.63 \\
  UGC5139       &  27.95 &  -13.09 &  -13.26 & ( -14.04) & ( -14.44) & ( -14.66) & ( -14.70) & \nodata &  -14.07 &  -13.53 &  \nodata &  -13.24 \\
  UGC5336       &  27.79 &  -12.79 &  -12.92 &  -13.39 &  -13.86 &  -14.24 &  -14.41 &  -14.39 &  -13.60 &  -13.30 &  \nodata &  -14.65 \\
\enddata
\tablecomments{All photometry is given as AB absolute magnitudes and corrected for Milky Way extinction.}
\tablenotetext{a}{Distance modulus.}
\tablenotetext{b}{Numbers within parentheses are for the Johnson-Cousins filters, numbers without parentheses are for the SDSS system. }
\end{deluxetable}

\begin{deluxetable}{lccc}
\tablecolumns{4}
\tablecaption{Average SED differences \label{tbl:offsets}}
\tablehead{\colhead{Filter} & \colhead{$\langle \log L_{mod}/L_{obs}\rangle$} & \colhead{$\sigma(\log L_{mod}/L_{obs})$} & \colhead{N$_{obs}$ } }
\startdata
$        FUV$ &   -0.009 &    0.125 & 38 \\
$        NUV$ &   -0.028 &    0.145 & 42 \\
$          u$ &   -0.104 &    0.099 & 21 \\
$          U$ &   -0.020 &    0.214 & 15 \\
$          B$ &   -0.005 &    0.115 & 19 \\
$          g$ &   -0.009 &    0.066 & 21 \\
$          V$ &   -0.005 &    0.127 & 15 \\
$          r$ &    0.024 &    0.064 & 21 \\
$          R$ &    0.042 &    0.213 & 18 \\
$          i$ &    0.045 &    0.080 & 21 \\
$          z$ &    0.088 &    0.096 & 21 \\
$ 3.6\micron$ &    0.121 &    0.171 & 41 \\
$ 4.5\micron$ &    0.117 &    0.161 & 42 \\
$ 5.6\micron$ &    0.171 &    0.334 & 23 \\
$   8\micron$ &   -0.091 &    0.411 & 25 \\
\enddata
\end{deluxetable}

\begin{deluxetable*}{lccccc}
\tablecolumns{6}
\tablecaption{Modeled $\tau_{UV}$ as Function of $NUV-r$ Color\tablenotemark{a} \label{tbl:tau_fit}}
\tablehead{\colhead{Band}   &  \colhead{Flux Fraction} & \colhead{$a$}  &  \colhead{$b$} & \colhead{$\sigma(\log \tau_{UV})$} &  \colhead{$R$}}
\startdata
     FUV & 50\% & 6.567 & 0.506 & 0.417 & 0.700\\
     FUV & 80\% & 7.029 & 0.558 & 0.344 & 0.853\\
       NUV & 50\% & 6.453 & 0.687 & 0.342 & 0.831\\
       NUV & 80\% & 7.215 & 0.652 & 0.184 & 0.962\\
\enddata
\tablenotetext{a}{Fits are of the form $\log \tau_{Band,Fraction}=a+b(NUV-r)$.}
\end{deluxetable*}

\end{document}